\documentclass[twocolumn,fleqn,natbib]{svjour3}
\usepackage{graphicx,url}
\usepackage{amsmath,amsfonts,amssymb,amsopn,amscd}
\usepackage{mathptmx}
\usepackage[centerlast]{subfigure}
\usepackage{color}
\usepackage{colortbl}
\usepackage{epsfig}

\def\url#1{\textcolor{blue}{\underline{#1}}}	% URL in blue

\definecolor{violet}{rgb}{1.00,0.00,1.00}
\definecolor{turquoise}{rgb}{0.00,0.40,0.50}    % like "sea green"
\definecolor{lightred}{rgb}{0.9,0,0}		% light red
\definecolor{brickred}{rgb}{0.8,0.5,0}

\journalname{Journal of Computational Neuroscience}

\emergencystretch=\maxdimen
\begin{document}

%\renewcommand{\baselinestretch}{1.1} \small\large

% ------------------------------------------------------------
% Title page
% ------------------------------------------------------------

\title{Comparative power spectral analysis of simultaneous
elecroencephalographic and magnetoencephalographic recordings in
humans suggests non-resistive extracellular media}

% Short title (if needed)
% \subtitle{EEG and MEG power spectra}

\author{Nima Dehghani$^*$, Claude B\'edard$^*$, Sydney S.\ Cash, Eric
Halgren and Alain Destexhe}

\institute{Integrative and Computational Neuroscience Unit (UNIC), \\
UPR2191, CNRS, Gif-sur-Yvette, France; Multimodal Imaging Lab., Dept. Radiology
and Neurosciences, UCSD, La Jolla, USA; Dept. Neurology, MGH, Harvard, Boston,
USA \\ *: co-first authors \\
\email{Destexhe@unic.cnrs-gif.fr}}

\date{\today}

\maketitle

% ------------------------------------------------------------
% Abstract
% ------------------------------------------------------------
% Please provide an abstract of 100 to 150 words. The abstract should
% not contain any undefined abbreviations or unspecified references.

\begin{abstract}

The resistive or non-resistive nature of the extracellular space in
the brain is still debated, and is an important issue for correctly
modeling extracellular potentials.  Here, we first show
theoretically that if the medium is resistive, the frequency
scaling should be the same for electroencephalogram (EEG) and
magnetoencephalogram (MEG) signals at low frequencies ($<$10~Hz). 
To test this prediction, we analyzed the spectrum of simultaneous
EEG and MEG measurements in four human subjects.  The frequency
scaling of EEG displays coherent variations across the brain, in
general between $1/f$ and $1/f^2$, and tends to be smaller in
parietal/temporal regions. In a given region, although the
variability of the frequency scaling exponent was higher for MEG
compared to EEG, both signals consistently scale with a different
exponent.  {In some cases, the scaling was similar, but only
when the signal-to-noise ratio of the MEG was low.  Several methods
of noise correction for environmental and instrumental noise were
tested, and they all increased the difference between EEG and MEG
scaling.}  In conclusion, there is a significant difference in
frequency scaling between EEG and MEG, which can be explained if
the extracellular medium ({including other layers such as dura
matter and skull}) is globally non-resistive.

\end{abstract}

% ------------------------------------------------------------
% Key Words
% ------------------------------------------------------------
% Please provide 4 to 6 keywords which can be used for indexing


% ------------------------------------------------------------
% Introduction
% ------------------------------------------------------------

\section{Introduction}

An issue central to modeling local field potentials is whether the
extracellular space around neurons can be considered as a resistive
medium.  A resistive medium {is equivalent to replacing the
medium by a simple resistance, which} considerably simplifies the
computation of local field potentials, as the equations to
calculate extracellular fields are very simple and based on
Coulomb's law (Rall and Shepherd, 1968; Nunez and Srinivasan,
2005).  {Forward models of the EEG and inverse
solution/source localization methods also assume that the medium is
resistive (Sarvas, 1987; Wolters and de Munck, 2007; Ramirez,
2008).  However, if the medium is non-resistive, the equations
governing the extracellular potential can be considerably more
complex because the quasi-static approximation of Maxwell equations
cannot be made (B\'edard et al., 2004).}

Experimental characterizations of extracellular resistivity are
contradictory.  Some experiments reported that the conductivity is
strongly frequency dependent, and thus that the medium is
non-resistive (Ranck, 1963; Gabriel et al., 1996a, 1996b, 1996c). 
Other experiments reported that the medium was essentially
resistive (Logothetis et al., 2007).  However, both types of
measurements used current intensities far larger than physiological
currents, which can mask the filtering properties of the tissue by
preventing phenomena such as ionic diffusion (B\'edard and
Destexhe, 2009).  Unfortunately, the issue is still open because
there exists no measurements to date using (weak) current
intensities that would be more compatible with biological current
sources.

In the present paper, we propose an indirect method to estimate if
extracellular space can be considered as a purely resistive medium.
We start from Maxwell equations and show that if the medium was
resistive, the frequency-scaling of electroencephalogram (EEG) and
magnetoencephalogram (MEG) recordings should be the same.  We then
test this scaling on simultaneous EEG and MEG measurements in
humans.

% By showing different frequency scaling, these results validate
% the assumption of non-resistivity

% ------------------------------------------------------------
% Methods
% ------------------------------------------------------------
\section{Methods}

\subsection{Participants and MEG/EEG recordings}

We recorded the electromagnetic field of the brain during quiet
wakefulness (with alpha rhythm occasionally present) from four
healthy adults (4 males ages 20-35).  Participants had no
neurological problems including sleep disorders, epilepsy, or
substance dependence, were taking no medications and did not
consume caffeine or alcohol on the day of the recording. We used a
whole-head MEG scanner (Neuromag Elekta) within a magnetically
shielded room (IMEDCO, Hagendorf, Switzerland) and recorded
simultaneously with 60 channels of EEG and 306 MEG channels
(Nenonen et al., 2004). MEG SQUID (super conducting quantum
interference device) sensors are arranged as triplets at 102
locations; each location contains one ``magnetometer'' and two
orthogonal planar ``gradiometers'' (GRAD1, GRAD2). Unless otherwise
noted, MEG will be used here to refer to the magnetometer
recordings. Locations of the EEG electrodes on the scalp of
individual subjects were recorded using a 3D digitizer (Polhemus
FastTrack). HPI (head position index) coils were used to measure
the spatial relationship between the head and scanner. Electrode
arrangements were constructed from the projection of 3D position of
electrodes to a 2D plane in order to map the frequency scaling
exponent in a topographical manner.  All EEG recordings
were monopolar with a common reference. Sampling rate was 1000 Hz.

{For all subjects, four types of consecutive recordings were
obtained, in the following order: (1) Empty-room recording; (2)
Awake ``idle'' recording where subjects were asked to stay
comfortable, without movements in the scanner, and not to focus on
anything specific; (3) a visual task; (4) sleep recordings. All
idle recordings used here were made in awake subjects with eyes
open, where the EEG was desynchronized.}  A few minutes of such
idle time was recorded in the scanner. For each subject, 3 awake
segments with duration of 60 seconds were selected from the idle
recordings (see example signals in Fig.~\ref{EEGMEG}).

As electrocardiogram (ECG) noise often contaminates MEG recordings,
Independent component analysis (ICA) algorithm was used to remove
such contamination; either Infomax (Bell and Sejnowski, 1995) or
the ``Jade algorithm'' from the EEGLAB toolbox (Delorme and Makeig,
2004) was used to achieve proper decontamination. In all
recordings, the ECG component stood out very robustly. In order not
to impose any change in the frequency content of the signal, we did
not use the ICA to filter the data on any prominent independent
oscillatory component and it was solely used to decontaminate the
ECG noise. We verified that the removal of ECG did not change the
scaling exponent (not shown).

In each recording session, just prior to brain recordings, we
recorded a few minutes of the electromagnetic field present within
the dewar in the magnetic shielded room.  Similar to wake epochs, 3
segments of 60 seconds duration were selected for each of the four
recordings.  This will be referred to ``empty room''
recordings and will be used in noise correction of the awake
recordings.

In each subject, the power spectral density (PSD) was
calculated by first computing the Fast Fourier transform (FFT) of 3
awake epochs, then averaging their respective PSDs (square modulus
of the FFT).  This averaged PSD was computed for all EEG and MEG
channels in order to reduce the effects of spurious peaks due to
random fluctuations.  The same procedure was also followed for
empty-room signals.

{\subsection{Noise correction methods}}

{Because the environmental and instrumental sources of noise
are potentially high in MEG recordings, we took advantage of the
availability of empty-room recordings to correct for the presence
of noise in the signal.  We used five different methods for noise
correction, based on different assumptions about the nature of the
noise.  We describe below these different correction methods, while
all the details are given in {\it Supplementary Methods}.}

{A first procedure for noise correction, exponent subtraction
(ES), assumes that the noise is intrinsic to the SQUID sensors.
This is justified by the fact that the frequency scaling of some of
the channels is identical to that of the corresponding empty-room
recording (see Results).  In such a case, the scaling is assumed to
entirely result from the ``filtering'' of the sensor, and thus the
correction amounts to subtract the scaling exponents.}

{A second class of noise subtraction methods assume that the
noise is of ambient nature and is uncorrelated with the signal. 
This chatacteristics, warrants the use of spectral subtraction
(where one subtracts the PSD of the empty-room from that of the MEG
recordings), prior to the calculation of the scaling exponent.  The
simplest form of spectral subtraction, linear multiband spectral
subtraction (LMSS), treats the sensors individually and does not
use any spatial/frequency-based statistics in its methodology (Boll
et al., 1979). An improved version, nonlinear multiband spectral
subtraction (NMSS), takes into account the signal-to-noise ratio
(SNR) and its spatial and frequency characteristics (Kamath and
Loizou, 2002; Loizou, 2007).  A third type, Wiener filtering (WF),
uses a similar approach as the latter, but obtain an estimate of
the noiseless signal from that of the noisy measurement through
minimizing the Mean Square Error (MSE) between the desired and the
measured signal (Lim et al., 1979; Abd El-Fattah et al., 2008).}

{A third type of noise subtraction, partial least squares
(PLS) regression, combines Principal component analysis (PCA)
methods with multiple linear regression (Abdi, 2010; Garthwaite,
1994).  This methods finds the spectral patterns that are common in
the MEG and the empty-room noise, and removes these patterns from
the PSD.}

\subsection{Frequency scaling exponent estimation}

{The method to estimate the frequency scaling exponent was
composed of  steps: First, applying a spline to obtain a smooth
FFT without losing the resolution (as can happen by using other
spectral estimation methods)}; Second, using a simple polynomial
fit to obtain the scaling exponent.  To improve the slope
estimation, we approximated the PSD data points using a spline,
which is a series of piecewise polynomials with smooth transitions
and where the break points (``knots'') are specified.  We used the
so-called ``B-spline'' (see details in de Boor, 2001).

The knots were first defined as linearly related to logarithm of
the frequency, which naturally gives more resolution to low
frequencies, to which our theory applies.  Next, in each frequency
window (between consecutive knots), we find the closest PSD value
to the mean PSD of that window. Then we use the corresponding
frequency as the optimized knot in that frequency range, leading
the final values of the knots.  The resulting knots stay close to
the initial distribution of frequency knots but are modified based
on each sensor's PSD data to provide the optimal knot points for
that given sensor (Fig.~\ref{knots}A).  We also use additional
knots at the outer edges of the signal to avoid boundary effects
(Eilers and Marx, 1996).  The applied method provides a reliable
and automated approach that uses our enforced initial frequency
segments with a high emphasis in low frequency and it optimizes
itself based on the data. After obtaining a smooth B-spline curve,
a simple 1st degree polynomial fit was used to estimate the slope
of the curve between 0.1-10 Hz {(the fit was limited to
this frequency band in order to avoid the possible effects of the
visible peak at 10~Hz on the estimated exponent}).Using this method provides a
reliable and robust estimate of the
slope of the PSD in logarithmic scale, as shown in Fig.~\ref{knots}B.  For more
details on the issue of automatic non-parametric fitting, and the rationale
behind combining the polynomial
with spline basis functions, we refer the reader to Magee, 1998
 as well as Royston \& Altman, 1994 and Katkovnik et al, 2006.

This procedure was realized on all channels automatically (102
channels for MEG, 60 channels for EEG, for each patient). 
{Every single fit was further visually confirmed.} In the case
of MEG, noise correction is essential {to validate the
results}.  For doing so, we used {different methods (as
described above) to reduce the noise}.  Next, all the mentioned
steps of frequency scaling exponents were carried out on the
corrected PSD. Results are shown in Fig.~\ref{topo}.

\subsection{Region of Interest (ROI)}

Three ROIs were selected for statistical comparisons of the
topographic plots. As shown in Figure~\ref{topo} (panel F), FR
(Frontal) ROI refers to the frontal ellipsoid, VX (Vertex) ROI
refers to the central disk located on vertex and PT
(Parietotemporal) refers to the horseshoe ROI.

% ------------------------------------------------------------
%  Results: theory section
% ------------------------------------------------------------

\section{Theory}

We start from first principles (Maxwell equations) and derive
equations to describe EEG and MEG signals.  Note that the
formalism we present here is different than the one usually given
(as in Plonsey, 1969; Gulrajani, 1998), because the linking
equations are here considered in their most general expression
(convolution integrals), in the case of a linear medium (see
Eq.~77.4 in Landau and Lifchitz, 1984).  This generality is
essential for the problem we treat here, because our aim is to
compare EEG and MEG signals with the predictions from the theory,
and thus the theory must be as general as possible.

\subsection{General formalism}

Maxwell equations can be written as
\begin{equation}
 \begin{array}{ccc}
~~\nabla\cdot\vec{D} = \rho^{free}~&~~&\nabla\cdot\vec{B}=0\\
\nabla\times\vec{E} = -\frac{\partial \vec{B}}{\partial
t}~&~&~~~~~~\nabla\times\vec{H}=\vec{j}+\frac{\partial \vec{D}}{\partial t}
\end{array}
\label{Max}
\end{equation}

If we suppose that the brain is linear in the electromagnetic sense
(which is most likely), then we have the two following linking
equations.  The first equation links the electric displacement with
the electric field:

\begin{equation}
 \vec{D}= \int_{-\infty}^{+\infty}\epsilon (\tau)\vec{E}(t-\tau )d\tau
\label{liaisonD}
\end{equation}
where $\mathbf{\epsilon }$ is a symmetric second-order tensor.

A second equation links magnetic induction and the magnetic
field:
\begin{equation}
 \vec{B}= \int_{-\infty}^{+\infty}\mu(\tau)\vec{H}(t-\tau )d\tau
\label{liaisonB}
\end{equation}
where $\mu$ is a symmetric second-order tensor.

If we neglect non-resistive effects such as diffusion (B\'edard and Destexhe,
2009), as well as any other nonlinear effects\footnote{Examples of nonlinear
  effects are variations of the macroscopic conductivity $\sigma_f$ with the
  magnitude of electric field $\vec{E}$.  {Such variations could appear
    due to ephaptic (electric-field) interactions for example.}  In addition,
  any type of linear reactivity of the medium to {the electric field or
    magnetic induction} can lead to frequency-dependent electric parameters
  $\sigma,\epsilon,\mu$ (for a detailed discussion of such effects, see
  B\'edard and Destexhe, 2009).}, then we can assume that the medium is
linear.  In this case, we can write:
\begin{equation}
 \vec{j}= \int_{-\infty}^{+\infty}\sigma(\tau)\vec{E}(t-\tau )d\tau
\label{liaisonj}
\end{equation}
where $\sigma$ is a symmetric second-order tensor
\footnote{Note that in textbooks, these linking equations
(Eqs.~\ref{liaisonD}--\ref{liaisonj}) are often algebraic and
independent of time (for example, see Eqs.~5.2-6, 5.2-7 and 5.2-8
in Gulrajani, 1998).  The present formulation is more general, more
in the line of Landau and Lifchitz (1984).}.  Because the effect
of electric induction (Faraday's law) is negligible, we can write:
{ 
\begin{equation} 
\begin{array}{ccccccc} 
\nabla\cdot\vec{D} &=& \rho^{free} & ~&\nabla\cdot\vec{B}&=&0\\
\nabla\times\vec{E} &=& 0
&~&\nabla\times\vec{H}&=&\vec{j}+\frac{\partial \vec{D}}{\partial
t} \end{array} \label{FF} 
\end{equation}} 
This system is much simpler compared to above, because electric
field and magnetic induction are decoupled.

By taking the Fourier transform of Maxwell equations (Eqs.~\ref{Max})
and of the linking equations
(Eqs.~\ref{liaisonD},\ref{liaisonB},\ref{liaisonj}), we obtain:
\begin{equation}
 \begin{array}{ccccccc}
\nabla\cdot\vec{D}_{f} & = &
 \rho_{f}^{free} & ~ &
 \nabla\cdot\vec{B}_{f} & = &  0
\\
\nabla\times\vec{E}_{f} & = & 0 & ~ &
\nabla\times\vec{H}_{f} & = &
\vec{j}_{f}+i\omega\vec{D}_{f}
\end{array}
\label{MaxFourier}
\end{equation}
where {$\omega=2\pi f$} and
\begin{equation}
 \begin{array}{ccc}
\vec{D}_{f} & = & \epsilon_{f}\vec{E}_{f} \\
\vec{B}_{f} & = &\mu_{f}\vec{H}_{f} \\
 \vec{j}_{f} &=&\vec{j}_{f}^{p}+\sigma_{f}\vec{E}_{f}
 \end{array}
\label{liaisonFourier}
\end{equation}
where the relation $\sigma_{f}\vec{E}_{f}$ in
Eq.~\ref{liaisonFourier} is the current density produced {by the
(primary) current sources in the extracellular medium. Note
that in this formulation, the electromagnetic parameters
$\epsilon_f$, $\mu_f$ and $\sigma_f$ depend on
frequency\footnote{In textbooks, the electric parameters are
sometimes considered as complex numbers, for example with the
notion of phasor (see Section~5.3 in Gulrajani, 1998), but they are
usually considered frequency independent.}.  This generalization
is essential if we want the formalism to be valid for media that
are linear but non-resistive, which can expressed with
frequency-dependent electric parameters.  It is also consistent
with the Kramers-Kronig relations (see Landau and Lifchitz, 1984;
Foster and Schwan, 1989).

$\vec{j}_{f }^p$ is the current density of these sources in Fourier
frequency space.  This current density is composed of the axial
current in dendrites and axons, as well as the transmembrane
current.} Of course, this expression is such that at any given
point, there is only one of these two terms which is non-zero. 
This is a way of preserving the linearity of Maxwell equations. 
Such a procedure is legitimate because the sources are not affected
by the field they produce\footnote{If it was not the case, then the
source terms would be a function of the produced field, which would
result in more complicated equations}.

\subsection{Expression for the electric field}

{From Eq.~\ref{MaxFourier}~{(Faraday's law in Fourier
space}), we can write:}
{
\begin{equation}
 \vec{E}_{f} = -\nabla V_{f} ~ .
\end{equation}}
{From Eq.~\ref{MaxFourier}~(Amp\`ere-Maxwell's law in Fourier space),
  we can write:}
{
\begin{eqnarray}
\nabla\cdot(\nabla\times\vec{H}_{f})
& = & \nabla\cdot\vec{j}_{f}
+i\omega\nabla\cdot(\epsilon_{f}\vec{E}_{f}) \nonumber \\
& = & \nabla\cdot\vec{j}_{f}^p - \nabla\cdot((\sigma_{f}
+i\omega\epsilon_{f})\nabla V_{f})=0
\end{eqnarray}}
{Setting $\gamma_{f} = \sigma_{f}+i\omega\epsilon_{f}$, one obtains:}
{
\begin{equation}
 \nabla\cdot(\gamma_{f}\nabla V_{f})=\nabla\cdot\vec{j}_{f}^p
\label{fondement1}
\end{equation}}
where {$\nabla\cdot\vec{j}_{f}^p $} is a source term and
$\mathbf{\gamma}_{f}$ is a symmetric second-order tensor ($3\times
3$). {Note that this tensor depends on position and frequency in general,
  and cannot be factorized.} We will call this expression
(Eq.~\ref{fondement1}) the ``first fundamental equation'' of the problem.

\subsection{Expression for magnetic induction}

{From the mathematical identity}
\begin{equation}
 \nabla\times\nabla\times\vec{X}=-\nabla^2\vec{X}+
\nabla(\nabla\cdot\vec{X})
\label{identite}
\end{equation}
it is clear that this is sufficient to know the divergence and the
curl of a field $\vec{X}$, because the solution of $\nabla^2{X}$ is
unique with adequate boundary conditions.

As in the case of magnetic induction, the divergence is necessarily
zero, it is sufficient to give an explicit expression of the curl as
a function of the sources.

Supposing that {$\mu=\mu_o\delta(t)$} is a scalar (tensor where all
directions are eigenvectors), and taking the curl of Eq.~\ref{MaxFourier} (D),
multiplied by the inverse of {$\gamma_f$}, we obtain the following
{equality}:
{
\begin{equation}
 \nabla\times(\gamma_f^{inv}\nabla\times\vec{B}_f)
=\mu_o \nabla\times(\gamma_f^{inv}\vec{j}_{f}^p)
\label{fondement2}
\end{equation}}
because {$\nabla\times\vec{E}_f=0$}.  This expression
(Eq.~\ref{fondement2}) will be named the ``second fundamental
equation''.

\subsection{Boundary conditions}

We consider the following boundary conditions:

1 - on the skull, we assume that {$V_f(\vec{r})$ is differentiable in
  space}, which is equivalent to assume that the electric field is finite.

2 - on the skull, we assume that {$\hat{n}\cdot\gamma_f\nabla V_f$} is
also continuous, which is equivalent to assume that the flow of current is
continuous.  Thus, we are interested in solutions where the electric field is
continuous.

3 - because the current is zero outside of the head, the current
perpendicular to the surface of cortex must be zero as well.  Thus,
the projection of the current on the vector $\hat{n}$ normal to the
skull's surface, must also be zero.
{
\begin{equation}
 \hat{n}(\vec{x})\cdot\gamma_f\nabla V_f (\vec{x}) =0
\end{equation}}
The latter expression can be proven by calculating the total current
and apply the divergence theorem (not shown).

\subsection{Quasi-static approximation to calculate magnetic induction}

The ``second fundamental equation'' above implies inverting {$\gamma_f$},
which is not possible in general, because it would require prior knowledge of
both conductivity and permittivity in each point outside of the sources.  If
the medium is purely resistive {($\gamma_f=\gamma $ where $\gamma$ is
  independent of space and frequency)}, one can evaluate the electric field
first, and next integrate {$\vec{B}_f$} using the quasi-static
approximation (Amp\`ere-Maxwell's law).  Because for low frequencies, we have
necessarily {$\vec{j}_f>>i\omega\vec{D}_f$}, we obtain {
$$\nabla\times\vec{B}_f=\mu_o\vec{j}_f ~ , $$ }
which is also known as Amp\`ere's law in {Fourier space}.

Thus, for low frequencies, one can skip the second fundamental equation.  Note
that in case this quasi-static approximation cannot be made (such as for high
frequencies), then one needs to solve the full system using both fundamental
equations.  {Such high frequencies are, however, well beyond the
  physiological range, so for EEG and MEG signals, the quasi-static
  approximation holds if the extracellular medium is resistive, or more
  generally if the medium satisfies $\nabla\times\vec{E}_f=-i\omega\vec{B}_f
  \backsimeq 0$ (see Eqs.~\ref{FF} and \ref{MaxFourier}).}

According to the quasi-static approximation, and using the linking
equation between current density and the electric field
(Eq.~\ref{liaisonFourier}), we can write:
{
\begin{equation}
 \nabla\times\vec{B}_f = \mu_o(\vec{j}_f^p-\gamma\nabla V_f)
\end{equation}}

Because the divergence of magnetic induction is zero, we have from
Eq.~\ref{identite}:
{
\begin{equation}
 \nabla\times\nabla\times\vec{B}_f = -\nabla^2 \vec{B}_f=
-\mu_o\nabla\times (\vec{j}_f^p-\gamma\nabla V_f)
\end{equation}}

This equation can be easily integrated using Poisson integral ({``Poisson
  equation''} for each component in Cartesian coordinates) {In Fourier
  space, this integral is given by the following expression}
{
\begin{equation}
\vec{B}_f(\vec{r})=\frac{\mu_o}{4\pi}\iiint\limits_{head}
\frac{\nabla\times (\vec{j}_f^p(\vec{r'})-\gamma\nabla V_f (\vec{r'}))}{\|
\vec{r}-\vec{r'}\|}dv'
\label{solution2f}
\end{equation}}

\subsection{Consequences}

If the medium is purely resistive (``ohmic''), then $\gamma$ does not depend
on the spatial position {(see Bedard et al., 2004; Bedard and Destexhe,
  2009)} nor on frequency, so that the solution for the magnetic induction is
given by:
{
\begin{equation}
 \vec{B}_f(\vec{r})=\frac{\mu_o}{4\pi}\iiint\limits_{head}
\frac{\nabla\times \vec{j}_f^p(\vec{r})}{\| \vec{r}-\vec{r'}\|}dv'
\label{solution2a}
\end{equation}}
and does not depend on the nature of the medium.

{For the electric potential, from Eq.~\ref{fondement1}, we obtain the
solution:
\begin{equation}
 V_f(\vec{x})=-\frac{1}{4\pi\gamma}\iiint\limits_{head}
\frac{\nabla\cdot \vec{j}_f^p}{| \vec{x}-\vec{x'}|}dv'
\label{solution1a}
\end{equation}}

{Thus, when the two source terms $\nabla\times \vec{j}_f^p$ and
  $\nabla\cdot \vec{j}_{f}^p$ are white noise, the magnetic induction and
  electric field must have the same frequency dependence.  Moreover, because
  the spatial dimensions of the sources are very small (see appendices), we
  can suppose that the current density $\vec{j}_f^p(\vec{x})$ is given by a
  function of the form:
\begin{equation}
 \vec{j}_f^p(\vec{x}) = \vec{j}^{pe}(\vec{x})F(f)
\label{assump}
\end{equation}
such that $\nabla\times\vec{j}_f^p$ and $\nabla\cdot\vec{j}_f^p$ have the same
frequency dependence for low frequencies.  Eq.~\ref{assump} constitutes the
main assumption of this formalism.}

{In Appendix~A, we provide a more detailed justification of this
  assumption, based on the differential expressions of the electric field and
  magnetic induction in a dendritic cable.  Note that this assumption is most
  likely valid for states with low correlation such as desynchronized-EEG
  states or high-conductance states, and for low-frequencies, as we analyze
  here (see details in the appendices).}

{Thus, the main prediction of this formalism is that if the extracellular
  medium is resistive, then the PSD of the magnetic induction and of the
  electric potential must have the same frequency dependence.  In the next
  section, we will examine if this is the case for simultaneously recorded MEG
  and EEG signals.}

\section{Test on experimental data}

\begin{figure}[h!]
\centering
\includegraphics[width=\columnwidth]{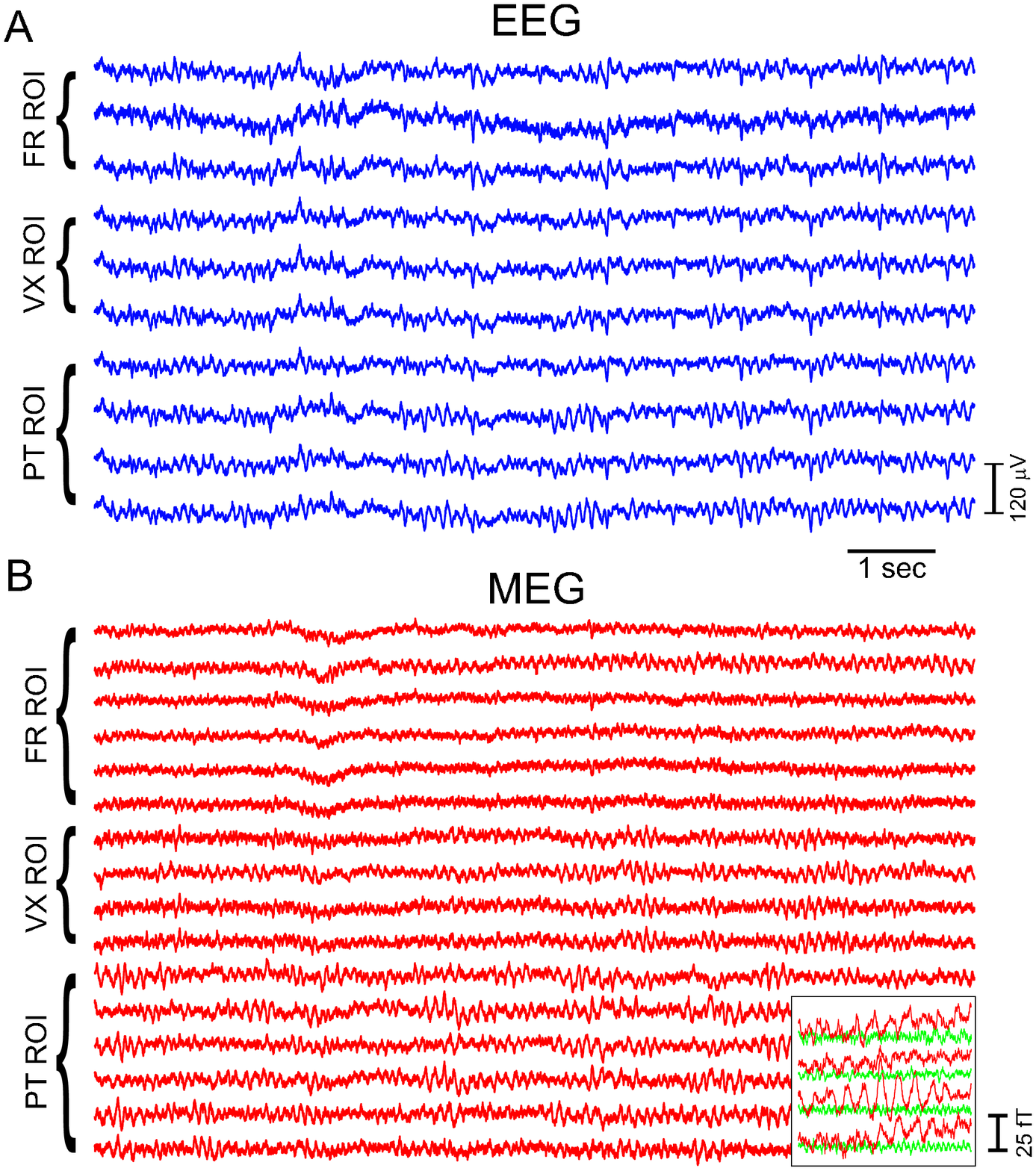}

\caption{Simultaneous EEG and MEG recordings in an awake human
subject.  This example shows a sample of channels from MEG/EEG
after ECG noise removal. Labels refer to ROIs as defined in methods
(also see Figure~\ref{topo}). FR: Frontal, VX:Vertex and PT:
Parietotemporal. These sample channels were selected to represent
both right and left hemispheres in a symmetrical fashion. {Inset:
magnification of the MEG (red) and ``empty-room'' (green) signals
superimposed from 4 sample channels.  All traces are before any
noise correction, but after ECG decontamination.}}

\label{EEGMEG}
\end{figure}

A total of 4 subjects were used for the analysis. 
Figure~\ref{EEGMEG} shows sample MEG and EEG channels from one of
the subjects, during quiet wakefulness. Although the subjects had
eyes open, a low-amplitude alpha rhythm was occasionally present
(as visible in Fig.~\ref{EEGMEG}).  There were also oscillations
present in the empty-room signal, but these oscillations are
evidently different from the alpha rhythm because of their low
amplitude and {the fact that} they do not appear in
gradiometers (see Suppl.\ Fig.~S1).

%-----------------------------------------------
%    Figure
%-----------------------------------------------
%-----------------------------------------------

In the next sections, we start by briefly presenting the method
that was used to estimate the frequency scaling of the PSDs. Then
we report the scaling exponents for 0.1-10~Hz frequency bands and
their differences in EEG and MEG recordings.

\subsection{Frequency scaling exponent estimation}

\begin{figure}[h!]
\centering
\includegraphics[width=0.8\columnwidth]{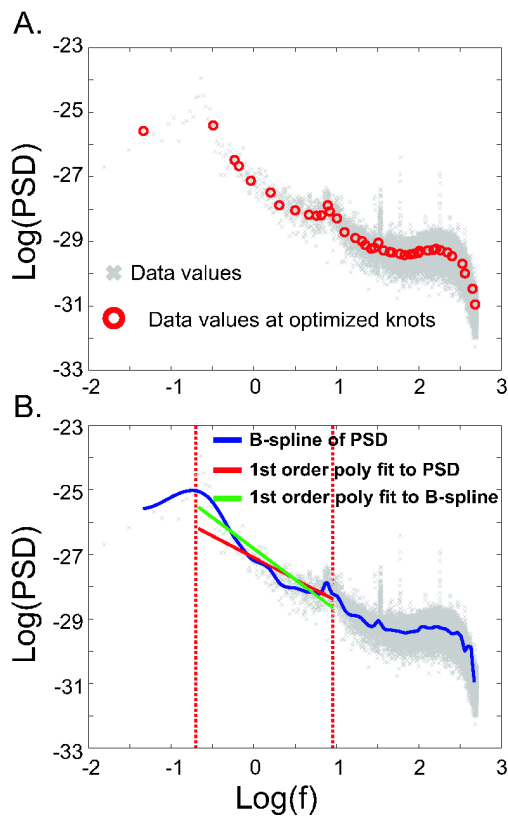}

\caption {A.log-log scale of the PSD vs frequency of a sample MEG
sensor along with the corresponding log(PSD) values (shown as
circles) at optimized knots in log-scale.  B. 1st degree Polynomial
fit on B-spline curve effectively captures properties of the signal
better than simple polynomial fit {and avoids the 10~Hz peak}. The fit was
limited between 0.1 to 10 Hz excluding the boundaries. This limits the fit
approximation to the next limiting optimized knots (between 0.1 and 0.2 to
between 9 and 10 Hz) to avoid the peaks at alpha and low frquencies (shown by
vertical dotted lines).}

\label{knots}
\end{figure}

Because of the large number of signals in the EEG and MEG
recordings, we used an automatic non-parametric procedure to
estimate the frequency scaling (see Methods). We used a B-spline
approximation by interpolation with boundary conditions to find a
curve which best represents the data(see Methods).  A high density
of knots was given to the low-frequency band (0.1-10~Hz), to have
an accurate representation of the PSD in this band, and calculate
the frequency scaling. An example of optimized knots to an
individual sensor is shown in Figure~\ref{knots}A; note that this
distribution of knots is specific to this particular sensor.  The
resulting B-spline curves were used to estimate the frequency
scaling exponent using a 1st degree polynomial fit. 
Figure~\ref{knots}B shows the result of the B-spline analysis with
optimized knots (in green) capturing the essence of the data better
than the usual approximation of the slope using polynomials (in
red).  The goodness of fit showed a robust estimation of the slope
using B-spline method.  Residuals were {-0.01 $\pm$ 0.6  for
empty-room, 0.2 $\pm$ 0.65 for MEG awake, 0.05 $\pm$ 0.6 for LMSS,
0.005 $\pm$ 0.64 for NMSS, 0.08 $\pm$ 0.5 for WF,0.001 $\pm$ 0.02
for PLS, and -0.02 $\pm$ 0.28 for EEG B-spline (all numbers to be
multiplied by 10$^{-14}$)}.

%-----------------------------------------------
%    Figure
%-----------------------------------------------
%-----------------------------------------------

\subsection{MEG and EEG have different frequency scaling exponents}

\begin{figure}[h!]
\centering
\includegraphics[width=1.05\columnwidth]{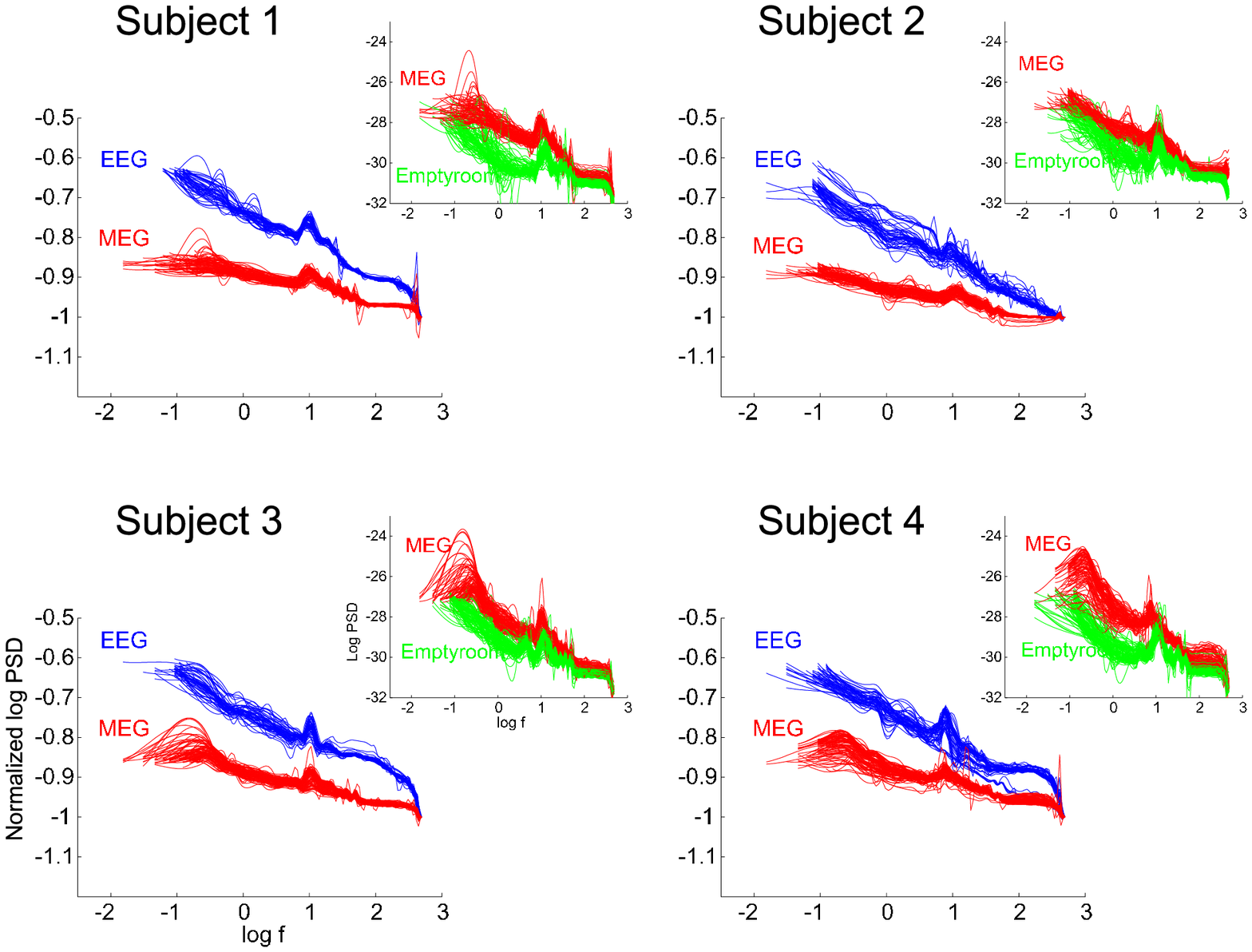}

\caption{B-spline fits of EEG awake and MEG awake {(prior to noise
correction)} recordings from all four subjects. Each line refers to
the fit of one sensor in log(PSD)-log(frequency) scale. For the
ease of visual comparison of the frequency scaling exponent,
log(PSD) values are normalized to their value at the maximum
frequency. Each panel represents the data related to one of our
four subjects. These plots show a clear distinction between the
frequency scaling of EEG and MEG. Insets show the comparison
between MEG awake {(prior to noise correction)} and MEG empty-room
recordings (not normalized).  Note that the empty-room scales the
same as the MEG signal, but in general EEG and MEG scale
differently.}

\label{PSD}
\end{figure}

Figure~\ref{PSD} shows the results of the B-spline curve fits to
the log-log PSD vs frequency for all sensors of all subjects. In
this figure, and only for the ease of visual comparison, these
curves were normalized to the value of the log(PSD) of the highest
frequency. As can be appreciated, all MEG sensors (in red) show a
different slope than that of the EEG sensors (in blue). The
frequency scaling exponent of the EEG is close to 1 ($1/f$
scaling), while MEG seems to scale differently. Thus, this
representation already shows clear differences of scaling between
EEG and MEG signals.

%-----------------------------------------------
%    Figure
%-----------------------------------------------
%-----------------------------------------------

However, MEG signals may be affected by ambient or instrumental
noise.  To check for this, we have analyzed the empty-room signals
using the same representation and techniques as for MEG, amd the
results are represented in Fig.~\ref{PSD} (insets). Empty-room
recordings always scale very closely to the MEG signal, and thus
the scaling observed in MEG may be due {in part} to environmental
noise or noise intrinsic to the detectors.  This emphasizes that it
is essential to use empty-room recordings made during the same
experiment to correct the frequency scaling exponent of MEG
recordings.

{To correct for this bias, we have used five different
procedures (see Methods).  The first class of procedure (ES)
considers that the scaling of the MEG is entirely due to filtering
by the sensors, which would explain the similar scaling between MEG
and empty-room recordings.  In this case, however, nearly all the
scaling would be abolished, and the corrected MEG signal would be
similar to white noise (scaling exponent close to zero).  Because
the similar scaling may be coincidental, we have used two other
classes of noise correction procedures to comply with different
assumptions about the nature of the noise.  The second class, is
composed of spectral subtraction (LMSS and NMSS) or Wiener
filtering (see Methods).  These methods are well-established in
other fields such as acoustics. The third class, uses statistical
patterns of noise to enhance PSD (PLS method, for details see
Methods).}

\subsection{Spatial variability of the frequency scaling exponent}

\begin{figure}[h!]
\centering
\includegraphics[width=\columnwidth]{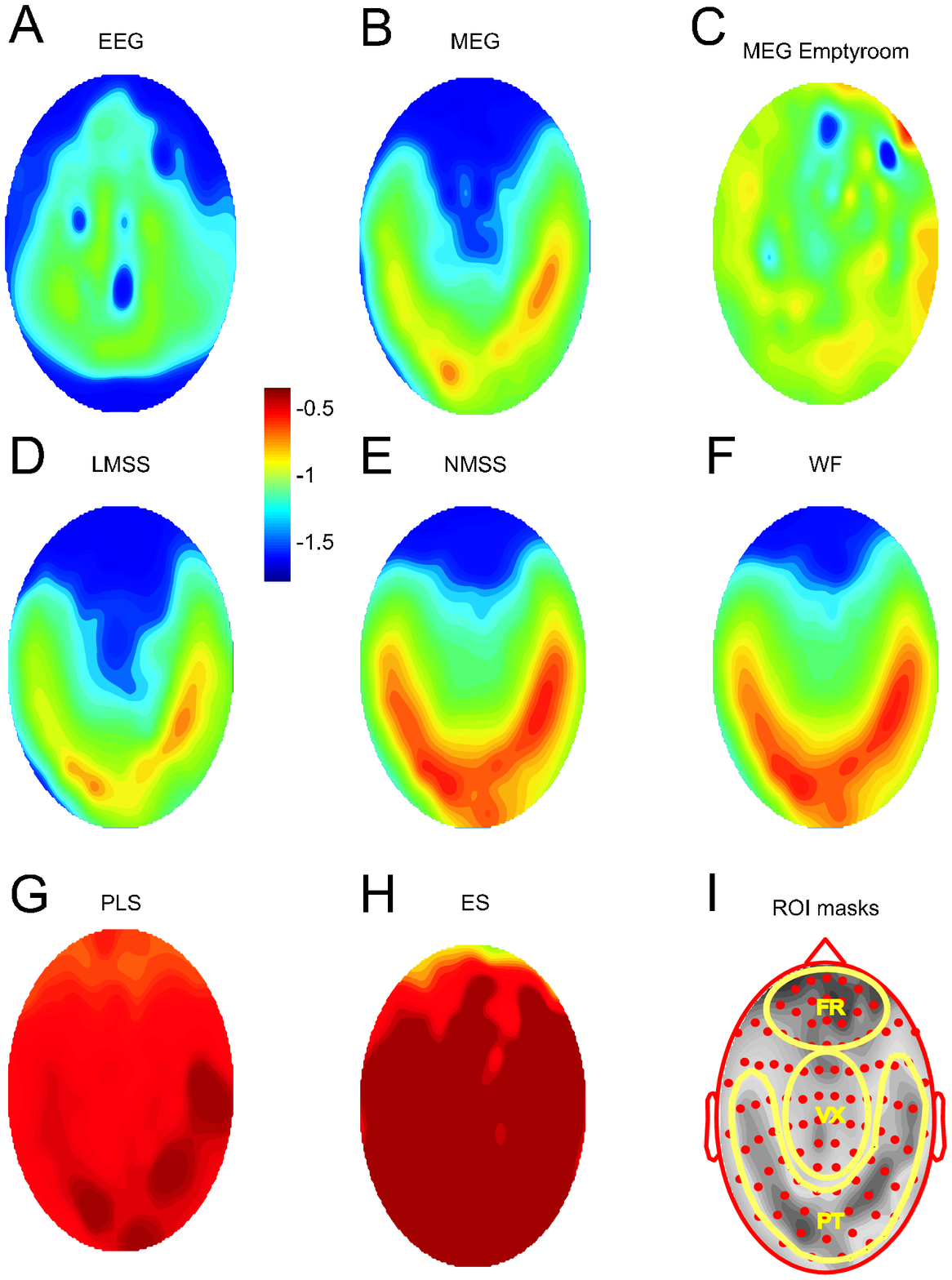}

\caption{Topographical representation of frequency scaling exponent
averaged across four subjects. A. EEG awake. B. MEG awake. C. MEG
empty-room.  {D, E. MEG after spectral subtraction of the
empty-room noise using linear (LMSS) and non-linear (NMSS) methods
respectively. F. MEG spectral enhancement using Wiener filtering
(WF). G. MEG, partial least square (PLS) approximation of non-noisy
spectrum. H. Exponent subtraction (the exponent represented is the
value of the frequency scaling exponent calculated for MEG signals,
subtracted from the scaling exponent calculated from the
corresponding emptyroom signals).  I. Spatial location of ROI masks
(shown in yellow). FR covers the Frontal, VX covers Vertex and PT
spans Parietotemporal.  Dots show spatial arrangement of 102 MEG
SQUID sensor triplets. The background gray-scale figure is same as
the one in panel B. Note that panels A through H use the same color
scaling. }}

\label{topo}
\end{figure}

We applied the above methods to all channels and represented the
scaling exponents in topographic plots in Fig.~\ref{topo}.  This
figure portrays that both MEG and EEG do not show a homogenous
pattern of the scaling exponent, confirming the differences of
scaling seen in Fig.~\ref{PSD}.  The EEG (Figure~\ref{topo}A) shows
that areas in the midline have values closer to 1, while those at
the margin can deviate from $1/f$ scaling. MEG on the other hand
shows higher values of the exponent in the frontal area and a
horseshoe pattern of low value exponents in parietotemporal regions
(Figure~\ref{topo}B).  {As anticipated above, empty-room
recordings scale more or less uniformly with values close to $1/f$
(Figure~\ref{topo}C), thus necessitating the correction for this
phenomena to estimate the correct MEG frequency scaling exponent. 
Different methods for noise reduction are shown in
Figure~\ref{topo}: spectral subtraction methods, such as LMSS
(Figure~\ref{topo}D), NMSS (Figure~\ref{topo}E), WF enhancement
(Figure~\ref{topo}F).  These corrections preserve the pattern seen
in Figure~\ref{topo}B, but tend to increase the difference with EEG
scaling: one method (LMSS) yields minimal correction while the
other two (NMSS and WF) use band-specific SNR information in order
to cancel the effects of background colored-noise (see Suppl.
Fig.~S2), and achieve higher degree of correction (see
Supplementary Methods for details).  Figure~\ref{topo}G portrays
the use of PLS to obtain a noiseless signal based on the noise
measurements.  The degree of correction achieved by this method is
higher than what is achieved by spectral subtraction and WF
methods. Exponent subtraction is shown in Figure~\ref{topo}H. This
correction supposes that the scaling is due to the frequency
response of the sensors, and nearly abolishes all the frequency
scaling (see also Suppl.\ Fig.~S3 for a comparison of different
methods of noise subtraction).}

%----------------------------------------------- 
% Figure
%----------------------------------------------- 
%-----------------------------------------------

\subsection{Statistical comparison of EEG and MEG frequency scaling}

Based on the patterns in Fig.~\ref{topo}, we created three ROIs
covering Vertex (FR), Vertex (VX) and the horseshoe pattern (PT).
These masks are shown in Fig.~\ref{topo}I.

%-----------------------------------------------
%    Figure
%-----------------------------------------------
\begin{figure}[h!]
\centering
\includegraphics[width=0.6\columnwidth]{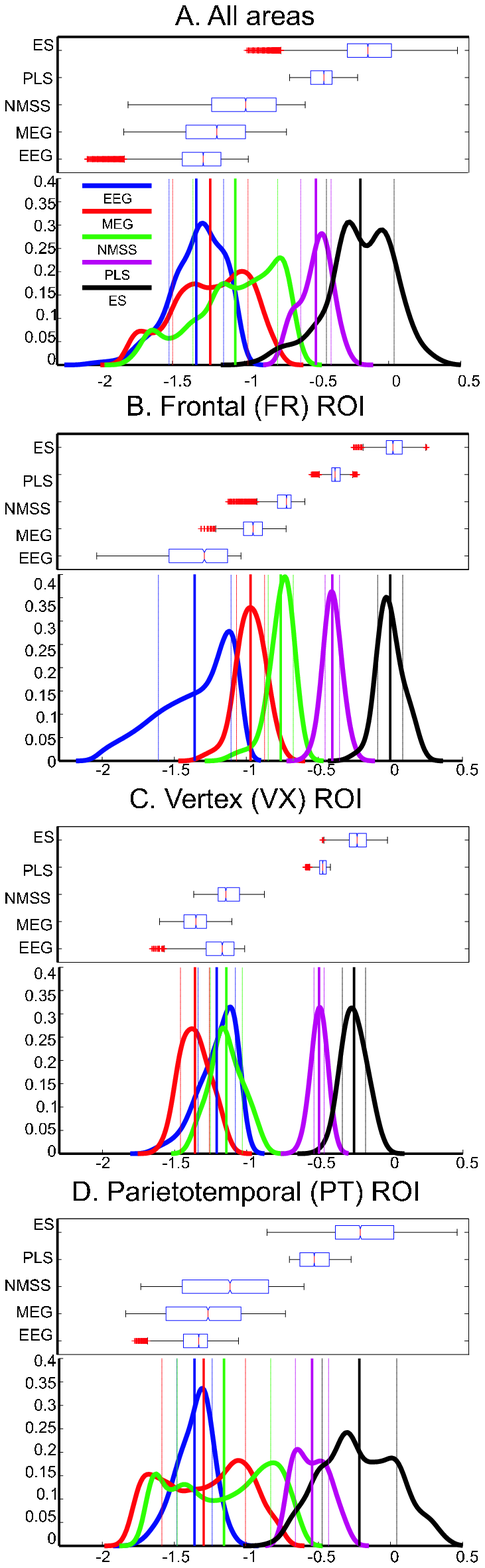}

\caption{Statistical comparison of EEG vs. MEG frequency scaling
exponent for all regions (A) and different ROI masks (B,C \& D). In
each panel, a box-plot on top is accompanied by a non-parametric
distribution function in the bottom. In the top graph, the box has
lines at the lower quartile, median (red), and upper quartile
values.  Smallest and biggest non-outlier observations (1.5 times
the interquartile range IRQ) are shown as whiskers.  Outliers are
data with values beyond the ends of the whiskers and are displayed
with a red + sign. In the bottom graph, a Non-parametric density
function shows the distribution of EEG, MEG and
empty-room-corrected MEG frequency scaling exponents {(note
that LMSS and WF are not shown here; see the text for
description.)}.  Thick and thin vertical lines show the mean and
mean $\pm$ std for each probability density function (pdf).}

\label{ROI}
\end{figure}
%-----------------------------------------------

Figure~\ref{ROI}A represents the overall pattern providing evidence
on the general difference and the wider variability in MEG
recordings. The next three panels relate to the {individual}
ROIs. {Of the spectral subtraction methods, NMSS achieves a
higher degree of correction in comparison with LMSS (see
Figure~\ref{topo}C, Figure~\ref{topo}D as well as Suppl. Fig.~S3). 
Because NMSS takes into account the effects of the background
colored-noise (Suppl. Fig.~S2), it is certainly more relevant to
the type of signals analyzed here. The results of NMSS and WF are
almost identical and confirm one another (see Figure~\ref{topo}E,
as well as Suppl.\ Fig.~S3). Therefore, of this family of noise
correction, only NMSS is portrayed here. Of the methods dealing
with different assumptions about the nature of the noise, the
``Exponent subtraction'' almost abolishes the frequency scaling
(Also see in Figure~\ref{topo}H, as well as Suppl.\ Fig.~S3). 
Applying PLS yields values in between ``Exponent subtraction'' and
that of NMSS and is portrayed in Figure~\ref{ROI}.}

{In the Frontal region (Figure~\ref{ROI}B), the EEG scaling
exponents show higher variance by comparison to MEG. Also, EEG
shows some overlaps with the distribution curve of non-corrected
MEG; this overlap becomes limited to the tail end of the NMSS
correction and is abolished in the case of PLS correction. As can
be appreciated, VX (Figure~\ref{ROI}C) shows both similar values
and similar distribution for EEG and non-corrected MEG.  These
similarities, in terms of regional overall values and distribution
curve, are further enhanced after NMSS correction.  It is to be
noted that, in contrast to these similarities, the one-to-one
correlation of NMSS and EEG at VX ROI are very low (see below,
Table~1B-C). The values of PLS noise correction are very different
from that of EEG and have a similar, but narrower, distribution
curve shape. Two other ROIs show distinctively different values and
distribution in comparing EEG and MEG.  Both NMSS and PLS agree on
this with PLS showing more extreme cases. Figure~\ref{ROI}D reveals
a bimodal distribution of MEG exponents in the parietotemporal
region (PT ROI).  This region has also the highest variance (in MEG
scaling exponents) compared to other ROIS. The distinction between
EEG and MEG is enhanced in PLS estimates; however, the variance of
PT is reduced in comparison to NMSS while the bimodality is still
preserved but weakened.}  The values of mean and standard deviation
for these ROIs' exponents are provided in Table~1A (mean $\pm$
standard deviation).

%-----------------------------------------------
%    Table
%-----------------------------------------------

\begin{table}[h!]

\ \\ {\bf A. Mean and standard deviation}

{
\begin{tabular}{c|lll}
       & EEG              & MEG (awake)          & NMSS\\
\hline
All    & -1.33 $\pm$ 0.19 & -1.24 $\pm$ 0.26 &  -1.06 $\pm$ 0.29 \\
FR ROI & -1.36 $\pm$ 0.25 & -0.97 $\pm$ 0.10 &  -0.76 $\pm$ 0.09 \\
VX ROI & -1.21 $\pm$ 0.13 & -1.36 $\pm$ 0.10 &  -1.14 $\pm$ 0.11 \\
PT ROI & -1.36 $\pm$ 0.12 & -1.30 $\pm$ 0.29 &  -1.16 $\pm$ 0.32 \\
\hline
\end{tabular}
}

\ \\ {\bf B. Pearson correlation}

{
\begin{tabular}{c|ll}
       & EEG vs. MEG & EEG vs. Corrected MEG (NMSS) \\
All    & 0.29 & 0.32 \\
FR ROI & 0.41 & 0.32 \\
VX ROI & -0.17 & -0.15 \\
PT ROI & 0.35 & 0.38 \\
\hline
\end{tabular}
}

\ \\ {\bf C. Kendall Rank Corr}

{
\begin{tabular}{c|ll}
       & EEG vs. MEG & EEG vs Corrected MEG (NMSS) \\
\hline
All    & 0.21 & 0.24 \\
FR ROI & 0.29 & 0.21 \\
VX ROI & -0.03 & -0.04 \\
PT ROI & 0.23 & 0.26 \\
\hline
\end{tabular}
}

\ \\ {\bf D. KruskalWallis}

{
\begin{tabular}{c|rlr}
 & p value & Chi-square & df Error \\
\hline
All                    & $<$ 10$^{-15}$ & 1.53 10$^{3}$  & 34838 \\
All noise-corrected    & $<$ 10$^{-15}$ & 8.03 10$^{3}$  & 34838 \\
FR ROI                 & $<$ 10$^{-15}$ & 3.30 10$^{3}$  & 5008 \\
FR ROI noise-corrected & $<$ 10$^{-15}$ & 3.72 10$^{3}$  & 5008 \\
VX ROI                 & $<$ 10$^{-15}$ & 1.72 10$^{3}$  & 5452 \\
VX ROI noise-corrected & $<$ 10$^{-15}$ & 0.23 10$^{3}$  & 5452 \\
PT ROI                 & $<$ 10$^{-15}$ & 0.21 10$^{3}$  & 13010 \\
PT ROI noise-corrected & $<$ 10$^{-15}$ & 1.18 10$^{3}$  & 13010 \\
\hline
\end{tabular}
}

\caption{ROI statistical comparison. A. mean and std of frequency
scale exponent for all regions and individual ROI. B. numerical
values of linear Pearson correlation. C. rank-based Kendall
correlation. D. non-parametric test of analysis of variance
(KruskalWallis). {Corrected MEG refers to spectral subtraction
using NMSS. The full table is provided in Supplementary
information.}}

\end{table}
%-----------------------------------------------

{The box-plots of Fig.~\ref{ROI}-plots further show the
difference between the medians, lower/upper quartile and
interquartile range.  The overall difference is that the
uncorrected MEG has much wider variance compared to EEG and
corrected MEG (in case of PLS correction); the absolute value of
the median of MEG (uncorrected, or corrected with either NMSS or
PLS) is always smaller than that of EEG. The VX region is an
exception to the above rules; interestingly, the one-to-one
correlation of VX happens to be the lowest of all (see below). In
the case of NMSS-corrected MEG, the shape of the pdf is preserved. 
However, PLS narrows the distribution curve of MEG but further
enhances the differences between MEG and EEG. Therefore, median and
lower/upper quartiles will have different value than that of EEG.}

Correlation values (Table~1B-C) show that, although VX ROI has the
closest similarity in terms of its central tendency and probability
distribution, it provides the lowest correlation in a pairwise
fashion. P-values (for testing the hypothesis of no correlation
against the alternative that there is a nonzero correlation) for
Pearson's correlation were calculated using a Student's
t-distribution for a transformation of the correlation and they
were all significant (less than 10$^{-15}$ for $\alpha$ = 0.05). 
Similarly, a non-parametric statistic Kendall tau rank correlation
was used to measure the degree of correspondence between two
rankings and assessing the significance of this correspondence
between MEG and EEG in the selected ROIs (Table~1C). P-values for
Kendall's tau and Spearman's rho calculate using the exact
permutation distributions were all significant (less than
10$^{-15}$ for $\alpha$ = 0.05).  Kendall tau shows that the rank
correlation for all areas considered together as well as for PT,
show a lesser correlation than that is shown by Pearson linear
correlation. Furthermore, we carried out a Kruskal-Wallis
nonparametric version of one-way analysis of variance.  We used
this test to avoid bias in ANOVA (KruskalWallis assumes that the
measurements come from a continuous distribution, but not
necessarily a normal distribution as is assumed in ANOVA). 
KruskalWallis uses analysis of variance on the ranks of the data
values, not the data values themselves and therefore is an
appropriate test for comparison of the homogeneity of pattern
between ROIs of two image as well as their statistical median. As
shown in Table~1D, all p-values were significant emphasizing the
difference between the spatial aspect of the spectral nature of MEG
and EEG. {Note that the difference of scaling exponent of EEG
and MEG was also confirmed by nonlinear spatial kendall correlation
analysis, independently of the ROIs classification (not shown).}

{\subsection{Relation of scaling exponent to signal-to-noise
ratio}}

{Noise correction does not affect all the sensors in a same
fashion. As presented in Suppl. Fig.~S3, the simple linear spectral
subtraction (LMSS) may lead to an increment or decrement of the
scaling exponent. In any case, the correction achieved by this
method is minimal. This is due to the fact that LMSS ignores the
complex non-linear patterns of the SNR in different channels
(Suppl.\ Fig.~S2). We show that for all subjects, as the frequency
goes up, the SNR goes down. It is also noticable that in each
defined frequency band, i.e. 0-10 Hz (Slow, Delta and Theta), 11-30
Hz (Beta), 30-80 Hz (Gamma), 80-200 Hz (Fast oscillation), 200-500
Hz (Ultra-fast oscillation), there is an observable
sensor-to-sensor SNR variability. This variability is at its
maximum in the band with the highest SNR (i.e. 1-10 Hz).  All
together, the non-linear nature of MEG SNR shows that a linear
spectral subtraction could behave non-optimally, leading to minimal
correction. This also conveys that the optimal spectral correction
can be achieved only by non-linear methods that explicitly take
into account the SNR information of the data. Therefore the
correction achieved by NMSS and WF have higher validity, in
agreement with the fact that both methods yield similar results in
terms of values and spatial distribution (Fig.~\ref{topo}E,
Fig.~\ref{topo}F).}

% ------------------------------------------------------------
%  Discussion
% ------------------------------------------------------------
%\clearpage

\section{Discussion}

In this paper, we have {used a combination of theoretical and
experimental analyses to investigate} the spectral structure of EEG
and MEG signals.  In the first part of the paper, we presented a
theoretical investigation showing that if the extracellular medium
is purely resistive, the equations of the frequency dependence of
electric field and magnetic induction take a simple form, because
the admittance tensor does not depend on spatial coordinates. 
Thus, the macroscopic magnetic induction does not depend on the
electric field outside the neuronal sources, but only depends on
currents inside neurons.  In this case, the frequency scaling of
the PSD should be the same for EEG and MEG signals.  This
conclusion is only valid in the linear regime, and for low
frequencies.

An assumption behind this formalism is that the spatial and
frequency dependence of the current density factorize
(Eq.~\ref{assump}).  We have shown in the appendices that this is
equivalent to consider the different current sources as
independent.  Thus, the formalism will best apply to states where
the activity of synapses is intense and of very low correlation. 
This is the case for desynchronized-EEG states or more generally
``high-conductance states'', in which the activity of neurons is
intense, of low correlation, and the neuronal membrane is dominated
by synaptic conductances (Destexhe et al., 2003).  In such
conditions, the dendrites are bombarded by intense synaptic inputs
which are essentially uncorrelated, and one can consider the
current sources as independent (Bedard et al., 2010).  In the
present paper, we analyzed EEG and MEG recordings in such
desynchronized states, where this formalism best applies.

Note that the above reasoning neglects the possible effect of
abrupt variations of impedances between different media (e.g.,
between dura matter and cerebrospinal fluid).  However, there is
evidence that this may not be influential.  First, our previous
modeling work (Bedard et al., 2004) showed that abrupt variations
of impedance have a negligible effect on low frequencies,
suggesting that even in the presence of such abrupt variations
should not play a role at low frequencies.  Second, in the
frequency range considered here, the skull and the skin are very
close to be resistive at low frequencies (Gabriel et al., 1996b),
so it is very unlikely that they play a role in the frequency
scaling in EEG and MEG power spectra even at high frequencies.

In the second part of the paper, we have analyzed simultaneous EEG
and MEG signals recorded in four healthy human subjects while awake
and eyes open (with desynchronized EEG). Because of the large
number of channels involved, we used an {automatic procedure
(B-splines analysis) to calculate the frequency scaling}.  As found
in previous studies (Pritchard, 1992; Freeman et al., 2000;
B\'edard et al., 2006a), we confirm here that the EEG displays
frequency scaling close to $1/f$ at low frequencies\footnote{{Note
that to compare scaling exponents between studies one must take
into account that the electrode montage may influence the scaling. 
For example, in bipolar (differential) EEG recordings, if two leads
are scaling as $1/(A+f)$ and $1/(B+f)$, the difference will have
regions scaling as $1/f^2$.}}.  However, this $1/f$ scaling was
most typical of the midline channels, while temporal and frontal
leads tended to scale with slightly larger exponents, up to $1/f^2$
(see Fig.~\ref{topo}A).  The same pattern was observed in all four
patients.

This approach differs from previous studies in two aspects. 
First, in contrast to prior studies (such as Novikov et al., 1997;
Linkenkaer-Hansen et al., 2001), we calculated the frequency
scaling of all the sensors and did not confine our analysis to a
specific region.  Second, unlike other investigators (such as Hwa
and Ferree, 2002a,b), we did not limit our evaluations to either
EEG or MEG alone, but rather analyzed the scaling of both type of
signals simultaneously. Such a strategy enables us to provide
an extended spatial analysis of the frequency scaling. It also
provides a chance to compare the scaling properties of these
signals in relation to their physical differences.

For the MEG recordings, the frequency scaling at low frequencies
was significantly lower compared to the EEG (see Fig.~\ref{PSD}). 
This difference in frequency scaling was also accompanied by
spatial variability patterns (see Fig.~\ref{topo}) showing three
distinct regions: 1) a frontal area where the exponents had their
highest values in the case of MEG; 2) a central area where the
values of exponents of EEG and MEG get closer to each other and 3)
a parietotemporal horseshoe region showing the lowest exponents for
MEG with bimodal characteristics (Fig.~\ref{ROI}).  In some cases,
the scaling of the uncorrected {and corrected} MEG signal was
also close to $1/f$, as reported previously (Novikov et al., 1997).
{In the frontal area (FR mask), the scaling exponent of the
EEG was generally larger. At Vertex (VX mask), EEG and MEG had
similar values and at the Parietotemporal region (PT mask), MEG
showed a bimodal property with a much broader range of scaling
exponent in comparison to EEG (see Fig.~\ref{topo}).} {Note that to
avoid the effect of spurious peaks, Novikov et al.\ used the
spectrum of signal differences and argued for the existence of a
local similarity regime in brain activity. This approach
fundamentally changes the spectral characteristics of Magnetometers
(which measure the absolute magnitude of the magnetic induction)
into a measure that only for the neighboring sensors approximates
the behavior of the gradiometers (which measures the gradient of
the magnetic induction). So it is not clear how to relate their
values to the ones obtained here.}

To make sure that the differences of frequency scaling between EEG
and MEG were not due to environmental or instrumental noise, we
have used {five different methods} to remove the effect of
noise.  {These methods are based on different assumptions
about the nature and effect of the noise}. A first possibility is
to correct for the noise induced by the MEG sensors.  It is known
that the SQUID detectors used in MEG recordings are very sensitive
to environmental noise and they can produce $1/f$ noise
(H\"am\"al\"ainen et al., 1993). {Under this assumption, part
of the scaling of the MEG could be due to ``filtering'' by the
sensors themselves, which justifies a simple subtraction of scaling
exponents to remove the effects of this filtering.}  Note that such
empty-room recordings were not possible for the EEG, although the
noise from the recording setup could be estimated (see Miller et
al., 2009 for example).  Because {in some cases} both MEG and
emptyroom signals have similar scaling, a simple correction by
subtracting the exponents would almost entirely abolish the
frequency scaling {while in other cases it may even revert the
sign of the scaling exponent (see Fig.~\ref{topo}H , Suppl.\
Fig.~S3).}

{However, if noise is not due to the sensors but is of
additive uncorrelated nature, then another method for noise
correction must be used.  For this reason, we have used a second
class of well-established methods consisting of spectral
subtraction (Boll et al.\ 1979; Sim et al., 1998).  Using three of
such methods (LMSS, NMSS and WF) changed the scaling exponent,
without fundamentally changing its spatial pattern
(Fig.~\ref{topo}D-F).  The largest correction was obtained by
non-linear methods which take into account the SNR information in
the MEG signal.  We also applied another class of method which uses
the collective characteristics of all frequencies in noise
correction (PLS).  Similar to exponent subtraction, this method
nearly abolished all the scaling of the MEG (Fig.~\ref{topo}G).  In
conclusion, although different methods for noise subtraction give
rise to different predictions about frequency scaling, all of the
used methods enhanced the difference between EEG and MEG scaling. 
Thus, we conclude that the difference of EEG and MEG scaling cannot
be attributed to noise, but is significant, therefore reinforcing
the conclusion that the medium must be non-resistive.}

An alternative method to investigate this is the ``Detrended
Fluctuation Analysis'' (DFA; see Kantelhardt et al., 2001;
Linkenkaer-Hansen et al., 2001; Hwa and Ferree, 2002a,b).  Like
many nonlinear approaches, DFA results are very vulnerable to the
selection of certain parameters. Different filters severely affect
the scaling properties of the electromagnetic signals to different
extents, and therefore the parameters estimated through the DFA
analysis could be false or lead to distorted interpretations of
real phenomena (Valencia et al., 2008), and these effects are
especially prominent for lower frequencies, which are precisely our
focus of investigation here. One of the fields for which DFA can
provide robust results is to analyze surrogate data with known
characteristics. Although the use of DFA to evaluate the scaling
exponents of EEG was vigorously criticized (Valencia et al., 2008),
a previous analysis (Hwa and Ferree, 2002a,b) reported two
different regions, a central and a more frontal, which somehow
correlate with the FR and VX regions identified in our analysis. 
Similarly, a study by Buiatti et al.\ (2007) using DFA
provided evidence for topographical differences in scaling
exponents of EEG recordings. They report that scaling exponents
were homogeneous over the posterior half of the scalp and became
more pronounced toward the frontal areas.  In contrast to
Linkenkaer-Hansen et al.\ (2004) (where envelope of alpha
oscillations was used for DFA estimation), this study uses the raw
signal in its DFA analysis and yields values closer to those
reported here.

Both uncorrected signals and empty-room correction show that there
is a fundamentally different frequency scaling between EEG and MEG
signals, with near-$1/f$ scaling in EEG, while MEG shows a wider
range at low frequencies.  Although it is possible that
non-neuronal sources affect the lower end ($<$1Hz) of the evaluated
frequency domain (Voipio et al., 2003), the solution to avoid these
possible effects remain limited to invasive methods such as
inserting the electrode into the scalp (Ferree et al., 2001) or
using intracranial EEG recordings (similar to Miller et al., 2009).
This approach would render wide range spatial recording as well as
simultaneous invasive EEG and MEG recordings technically demanding
or impractical. However, if technically feasible, such methods
could provide a way to bypass non-neuronal effects at very low
frequency. It could also provide a chance to evaluate the effects
of spatial correlation on spectral structure at a multiscale level.

The power spectral structure we observe here is consistent with a scenario
proposed previously (B\'edard et al., 2006a): the $1/f$ structure of the EEG
and LFP signals is essentially due to a frequency-filtering effect of the
signal through extracellular space; this type of scaling can be explained by
ionic diffusion and its associated Warburg
  impedance\footnote{Ionic diffusion can create an impedance known as
      the ``Warburg impedance'', which scales as $1/\sqrt{\omega}$, giving
      $1/f$ scaling in the power spectra (Taylor and Gileadi, 1995; Diard et
      al., 1999).} (see B\'edard and Destexhe, 2009).  It is also consistent
with the matching of LFPs with multi-unit extracellular recordings, which can
be reconciled only assuming a $1/f$ filter (B\'edard et al., 2006a).  Finally,
it is also consistent with the recent evidence from the transfer function
calculated from intracellular and LFP recordings, which also showed that the
extracellular medium is well described by a Warburg impedance (B\'edard et
al., 2010, submitted to this issue). If this non-resistive aspect of
  extracellular media is confirmed, it may influence the results of models of
  source localization, which may need to be reformulated by including more
  realistic extracellular impedances.

In conclusion, the present theoretical and experimental analysis
suggests that the scaling of EEG and MEG signals cannot be
reconciled using a resistive extracellular medium.  The $1/f$
structure of EEG with smaller scaling exponents for MEG
is consistent with non-resistive extracellular impedances,
such as capacitive media or diffusion (Warburg) impedances. 
Including such impedances in the formalism is non trivial because
these impedances are strongly frequency dependent.  The Poisson
integral (the solution of Poisson's law
$\nabla\cdot\vec{D}=-\nabla\cdot\epsilon\nabla V=\rho$) would not
apply anymore (see B\'edard et al., 2004; B\'edard and Destexhe,
2009). Work is under way to generalize the formalism and
include frequency-dependent impedances.

Finally, it is arguable that the scaling could also be influenced by the
  cancellation and the extent of spatial averaging of microscopic signals,
  which are different in EEG and MEG (for more details on cancellation see
  Ahlfors et al., 2010; for details on spatial sensitivity profile see Cuffin
  and Cohen, 1979).  Such a possible role of the complex geometrical
  arrangement of underlying current sources should be investigated by 3D
  models which could test specific assumptions about the geometry of the
  current sources and dipoles, and their possible effect on frequency
  scaling. Such a scenario constitutes another possible extension of the
  present study.

% ------------------------------------------------------------
%  Appendix
% ------------------------------------------------------------

\begin{appendix}

{\section*{Appendix}}

{\section{Frequency dependence of electric field and magnetic induction}}

{To compare the frequency dependence of magnetic induction and electric
  field, we evaluate them in a dendritic cable, expressed differentially.  For
  a differential element of dendrite, in Fourier space, the current produced
  by a magnetic field (Amp\`ere-Laplace law) is given by the following
  expression (see Appendix~B):
\begin{equation}
 \delta\vec{B}_f(\vec{r})=
 \frac{\mu_o}{4\pi} \vec{j}_f^p(\vec{r'})\times~\frac{\vec{r}-
\vec{r'}}{\|\vec{r}-\vec{r'}\|^3}~\delta v'
\label{difB}
\end{equation}
when the extracellular medium is resistive.  Note that the source of magnetic
induction is essentially given by the component of $\vec{j}_f^p$ along the
axial direction ($j_f^i$) within each differential element of dendrite because
the perpendicular (membrane) current does not participate to producing the
magnetic induction if we assume a cylindrical symmetry.}

{For the electric potential, we have the following differential
  expression for a resistive medium (see Appendix~C):
\begin{equation}
\delta\vec{V}_f(\vec{r})=\frac{1}{4\pi\gamma}\frac{ \delta
I_f^{\perp}(\vec{r}')}{\| \vec{r}-\vec{r'}\|}
=\frac{1}{4\pi\gamma}\frac{j_f^{m}(\vec{r}')}{\| \vec{r}-\vec{r'}\|}\delta S'
\label{difE}
 \end{equation}
where $j_f^{m}$ is the transmembrane current per unit of surface.}

{If we consider the differential expressions for the magnetic induction
  (Eq.~\ref{difB}) and electric potential (Eq.~\ref{difE}), one can see that
  the frequency dependence of the ratio of their modulus is completely
  determined by the frequency dependence of the ratio of current density
  $j_f^m$ and $j_f^i $.  In Appendix~D, we show that this ratio is
  quasi-independent of frequency for a resistive medium, for low frequencies
  (smaller that $\sim$10~Hz), and if the current sources are of very low
  correlation.}

{Thus, magnetic induction and electric potential can be very well
  approximated by:
\begin{small}\begin{equation}
\begin{array}{c}
V_f(\vec{r})=
N<V>=
N<\sum\limits_{l=1}^{N} \delta V_f^l>
 \\
\vec{B}_f(\vec{r})=
N<\vec{B}>=
N<\sum\limits_{l=1}^{N} \delta \vec{B}_f^l>
\end{array}
\end{equation}\end{small}
for sufficiently small differential dendritic elements ($N/l$ large).}

{Because the functions of spatial and frequency are statistically
  independent, we can write the following expressions for the square modulus
  of the fields (see Eqs.~\ref{difB} and \ref{difE}):
\begin{small}\begin{equation}
\begin{array}{ccccc}
|V_f(\vec{r})|^2 &=&
N^2|<\sum\limits_{l=1}^{N}V^l(\vec{r})G_l^m(f)>|^2 &=&
|V(\vec{r})|^2|G(f)|^2
 \\
\|\vec{B}_f(\vec{r})\|^2 &=&
N^2\|<\sum\limits_{i=1}^{N}\vec{B}^i(\vec{r})G_l^m(f)>\|^2 &=&
\|\vec{W}(\vec{r})\|^2|G(f)|^2
\end{array}
\end{equation}\end{small}}
{where \begin{small} $G(f)=<G_l^m(f)> $\end{small}, \begin{small}
    $V^l(\vec{r})=<V^l(\vec{r})>$ \end{small} and \begin{small}
    $\vec{W}(\vec{r})=<\vec{B}^l(\vec{r})>$ \end{small}. Thus, the scaling of
  the PSDs of the electric potential and magnetic induction must be the same
  for low frequencies (smaller than $\sim$10~Hz) if the medium is resistive
  and when the current sources have very low correlation.}

{\section{Differential expression for the magnetic induction}}\label{A1}

{According to Maxwell equations, the magnetic induction is given by:
\begin{equation}
 \vec{B}_f(\vec{r})=
 \frac{\mu_o}{4\pi}\iiint\limits_{head}
\frac{\nabla'\times\vec{j}_f^p(\vec{r'})}{\|\vec{r}-\vec{r'}
\|}~dv'
\end{equation}}
{where $dv'=dx^{'1}dx^{'2}dx^{'3}$ and
$$
\nabla'(\frac{1}{\|\vec{r}-\vec{r'}\|})=\frac{\vec{r}-
\vec{r'}}{\|\vec{r}-\vec{r'}\|^3}
$$
for a perfectly resistive medium.}

{We now show that this expression is equivalent to Ampere-Laplace law.}

{From the identity $\nabla' \times (g\vec{A})= g(\nabla' \times
  \vec{A})+\nabla' g\times\vec{A}$, where $\nabla' =
    \hat{e}_x\frac{\partial}{\partial x'}+\hat{e}_y\frac{\partial}{\partial
      y'}+\hat{e}_z\frac{\partial}{\partial z'}$, we can write:
\begin{equation}
 \vec{B}_f(\vec{r})=
 \frac{\mu_o}{4\pi}
 \iiint\limits_{head}
[\nabla'\times (\frac{\vec{j}_f^p(\vec{r'})}{\|\vec{r}-\vec{r'}\|})
+ \frac{\mu}{4\pi}
\vec{j}_f^p(\vec{r'})\times\nabla'\frac{1}{\|\vec{r}-\vec{r'}\|}]
~dv'
\end{equation}}
{Moreover, we also have the following identity
\begin{equation}
\iiint\limits_{head}
\nabla'\times (\frac{\vec{j}_f^p(\vec{r'})}{\|\vec{r}-\vec{r'}\|})~dv'
= -\iint\limits_{\partial head}
 \frac{\vec{j}_f^p(\vec{r'})}{\|\vec{r}-\vec{r'}\|}\times\hat{n}~dS'
\end{equation}}
{where $\hat{n}$ is a unitary vector perpendicular to the integration
  surface and going outwards from that surface.  Extending the volume integral
  outside the head, the surface integral is certainly zero because the current
  is zero outside of the head.  It follows that:}
{
\begin{equation}
 \vec{B}_f(\vec{r})=
 \frac{\mu_o}{4\pi}\iiint\limits_{head} 
\vec{j}_f^p(\vec{r'})\times~\frac{\vec{r}-
\vec{r'}}{\|\vec{r}-\vec{r'}\|^3}~dv'
\label{AmpeLapl}
\end{equation}}
{where $dv'=dx^{'1}dx^{'2}dx^{'3}$ because
$$
\nabla'(\frac{1}{\|\vec{r}-\vec{r'}\|})=\frac{\vec{r}-
\vec{r'}}{\|\vec{r}-\vec{r'}\|^3}
$$}
{Eq.~\ref{AmpeLapl} is called the Amp\`ere-Laplace law (see Eq.~13 in
  H\"am\"al\"ainen et al., 1993). It is important to note that this expression
  for the magnetic induction is not valid when the medium is not resistive.}

{Finally, from the last expression, the magnetic induction for a
  differential element of dendrite can be written as:}
{
\begin{equation}
 \delta\vec{B}_f(\vec{r})=
 \frac{\mu_o}{4\pi} \vec{j}_f^p(\vec{r'})\times~\frac{\vec{r}-
\vec{r'}}{\|\vec{r}-\vec{r'}\|^3}~\delta v'
\end{equation}
}

{\section{Differential expression of the electric field and electric
    potential}\label{A2}}

{In this appendix, we derive the differential expression for the electric
  field.  Starting from Eq.~\ref{fondement1}, we obtain the solution for the
  electric potential:}
{
\begin{equation}
 V_f(\vec{r})=-\frac{1}{4\pi\gamma_f}\iiint\limits_{head}
\frac{\nabla\cdot \vec{j}_f^p(\vec{r}')}{\Vert \vec{r}-\vec{r'}\Vert}~dv'
\label{solution1abis}
\end{equation}}

{It follows that the electric field produced by the ensemble of sources
  can be expressed as:} {
\begin{equation}
  \vec{E}_f(\vec{r})=-\nabla
V_f(\vec{r})=\frac{1}{4\pi\gamma_f}\iiint\limits_{head}
  \nabla\cdot \vec{j}_f^p(\vec{r}')\cdot\frac{ \vec{r}-\vec{r'}}{\|
\vec{r}-\vec{r'}\|^3}~dv'
\end{equation}}
{such that every differential element of dendrite produces the 
following electric field:}
{
\begin{equation}
 \delta\vec{E}_f(\vec{r})=
 \frac{\nabla\cdot \vec{j}_f^p(\vec{r}')}{4\pi\gamma_f}
\cdot
\frac{ \vec{r}-\vec{r'}}{\| \vec{r}-\vec{r'}\|^3}~
\delta v'
\end{equation}}

{The transmembrane current $\delta I_f^{\perp}$ obeys
  $\delta I_f^{\perp}=i\omega\rho_f(\vec{r}') \delta v'$ because we are in a
  quasi-stationary regime in a differential dendritic element.  Taking into
  account the differential law of charge conservation $\nabla\cdot
  \vec{j}_f(\vec{r}')=-i\omega\rho_f(\vec{r}')$, we have:}
{
\begin{equation}
\delta\vec{E}_f(\vec{r})=\frac{\delta I_f^{\perp}(\vec{r}')}{4\pi\gamma_f}\frac{
\vec{r}-\vec{r'}}{\| \vec{r}-\vec{r'}\|^3}
=\frac{j_f^{m}(\vec{r}')}{4\pi\gamma_f}\frac{ \vec{r}-\vec{r'}}{\|
\vec{r}-\vec{r'}\|^3}\delta S'
\end{equation}}
{where $j_f^{m}$ is the density of transmembrane current per unit surface
  and $\delta S'$ is the surface area of a differential dendritic element.
  This approximation is certainly valid for frequencies lower than 1000~Hz
  because the Maxwell-Wagner time (see Bedard et al., 2006b) of the cytoplasm
  ($\tau_{mw}^{cyto}=\epsilon/\sigma \sim 10^{-10}~s.$) is much smaller than
  the typical membrane time constant of a neuron ($\tau_m \sim 5-20~ms$).}

{Finally the contribution of a differential element of dendrite
to the electric potential at position $\vec{r}$ is given by}
{
\begin{equation}
\delta\vec{V}_f(\vec{r})=\frac{1}{4\pi\gamma_f}\frac{ \delta
I_f^{\perp}(\vec{r}')}{\| \vec{r}-\vec{r'}\|}
=\frac{1}{4\pi\gamma_f}\frac{ j_f^{m}(\vec{r}')}{\| \vec{r}-\vec{r'}\|}\delta S'
\end{equation}}

{We note that the expressions for the electric field and potential
  produced by each differential element of dendrite have the same frequency
  dependence because it is directly proportional to $\frac{j_f^m}{\gamma_f}$
  for the two expressions.  Also note that if the medium is resistive, then
  $\gamma_f=\gamma$ and the frequency dependence of the electric field and
  potential are solely determined by that of the transmembrane
  current $j_f^m$.}

{\section{Frequency dependence of the ratio
$ { j_f^i(\vec{x})} / {j_f^m(\vec{x})}$.}}\label{A3}

{For each differential element of dendrite, we consider the standard
  cable model, in which the impedance of the medium is usually neglected (it
  is usually considered negligible compared to the membrane impedance).  In
  this case, we have:}
{
\begin{equation}
\left\{
 \begin{array}{ccc}
 j_f^m &=&
 \frac{V_f^m}{r_m}+i\omega c_m V_f^m \\~\\
j_f^i &=& -\sigma\frac{\partial V_f^m}{\partial x} =
-\frac{1}{r_i}\frac{\partial V_f^m}{\partial x}
\end{array}
\right .
\label{cable}
\end{equation}}

{where $V_f^m$, $j_f^i$, $j_f^m$, $c_m$, $r_m$ et $r_i$ are respectively
  the membrane potential, the current density in the axial direction, the
  transmembrane current density, the specific capacitance ($F/m^2$), the
  specific membrane resistance ($\Omega.m^2$) and the cytoplasm resistivity
  ($\Omega.m$).}

{It follows that}
{
\begin{equation}
 \frac{ j_f^i(\vec{x})}{j_f^m(\vec{x})}
=\frac{r_m}{r_i(1+i\omega\tau_m)}
~\cdot\frac{\partial}{\partial x}[ln(V_f^m(\vec{x})]
\end{equation}}
{where $\tau_m=r_mc_m$.}

{Under {\it in vivo}--like conditions, the activity of neurons is
intense and of very low correlation.  This is the case for
desynchronized-EEG states, such as awake eyes-open
conditions, where the activity of neurons is characterized by very
low levels of correlations.  There is also evidence that in such
conditions, neurons are in ``high-conductance states'' (Destexhe et
al., 2003), in which the synaptic activity dominates the
conductance of the membrane and primes over intrinsic currents.  In
such conditions, we can assume that the synaptic current sources
are essentially uncorrelated and dominant, such that the
deterministic link between current sources will be small and can be
neglected (see Bedard et al., 2010). Further assuming that the
electric properties of extracellular medium are homogeneous, then
each differential element of dendrite can be considered as
independent and the voltages $V_m$ have similar power spectra.}

{In such conditions, we have:
\begin{equation}
 V_f^m (\vec{x}) = F^m(\vec{x})G^m(f)  \label{assump2}
\end{equation}
Note that this expression implies that we have in general for each
differential element of dendrite:
\begin{equation}
\left\{
\begin{array}{ccc}
 j_f^m (\vec{x}) &=& F^m(\vec{x})(\frac{1+i\omega\tau_m}{r_m})G^m(f)\\
~&~&~
\\
 j_f^i (\vec{x}) &=& -\frac{1}{r_i}\frac{\partial F^m(\vec{x})}{\partial
x}G^m(f)= F^i(\vec{x})G^m(f)
\end{array}
\right .
\end{equation}
according to Eq.~\ref{cable}.}

{It follows that
\begin{equation}
 \frac{ j_f^i(\vec{x})}{ j_f^m(\vec{x})}
=\frac{r_m}{r_i(1+i\omega\tau_m)}
~\cdot\frac{\partial}{\partial x}[ln(F(\vec{x}))]
\approx \frac{r_m}{r_i}
~\cdot\frac{\partial}{\partial x}[ln(F(\vec{x}))]
\end{equation}}

{Thus, for frequencies smaller than $1/(\omega \tau_m)$ (about 10 to
  30~Hz for $\tau_m$ of 5-20~ms), the ratio $ \frac{ j_f^i(\vec{x})}{
    j_f^m(\vec{x})}$ will be frequency independent, and for each differential
  element of dendrite, we have:
\begin{equation}
\left\{
\begin{array}{ccc}
 j_f^m (\vec{x}) &=& F^m(\vec{x})G^m(f)\\
 j_f^i (\vec{x}) &=& F^i(\vec{x})G^m(f)
\end{array}
\right .
\label{eqf}
\end{equation}
for frequencies smaller than $\sim$10~Hz.}
\end{appendix}

% ------------------------------------------------------------
%  Acknowledgments
% ------------------------------------------------------------

\section*{Acknowledgments}

{We thank Philip Louizo for comments on spectral subtraction
methods and Herv\'e Abdi for comments on Partial least square methods.} Research
supported by the Centre National de la Recherche Scientifique
(CNRS, France), Agence Nationale de la Recherche (ANR, France), the
Future and Emerging Technologies program (FET, European Union;
FACETS project) and the National Institutes of Health (NIH grants
NS18741, EB009282 and NS44623).  N.D. is supported by a fellowship
from Ecole de Neurosciences de Paris (ENP).  Additional information
is available at \url{http://cns.iaf.cnrs-gif.fr}

% ------------------------------------------------------------
%  References
% ------------------------------------------------------------

% Reference list entries should be alphabetized by the last names of
% the first author of each work.
%
% Journal article
% Harris, M., Karper, E., Stacks, G., Hoffman, D., DeNiro, R., Cruz,
% P., et al. (2001). Writing labs and the Hollywood connection.
% Journal of Film Writing, 44(3), 213–245.
%
% Book
% Calfee, R. C., & Valencia, R. R. (1991). APA guide to preparing
% manuscripts for journal publication. Washington, DC: American
% Psychological Association.
%
% Book chapter
% O'Neil, J. M., & Egan, J. (1992). Men's and women's gender role
% journeys: Metaphor for healing, transition, and transformation. In B.
% R. Wainrib (Ed.), Gender issues across the life cycle (pp. 107–123).
% New York: Springer.
%
% Online document
% Abou-Allaban, Y., Dell, M. L., Greenberg, W., Lomax, J., Peteet,
% J., Torres, M., Cowell, V. (2006). Religious/spiritual commitments
% and psychiatric practice. Resource document. American Psychiatric
% Association.
% http://www.psych.org/edu/other_res/lib_archives/archives/200604.pdf.
% Accessed 25 June 2007.

%\clearpage

\section*{References}

\begin{description}

\bibitem{ElFattah2008} {Abd El-Fattah MA, Dessouky MI, Diab SM
and Abd El-samie FE. (2008) Speech enhancement using an adaptive
Wiener filtering approach. {\it Prog. Electromagnetics Res.} {\bf 4}:
167-184.}

\bibitem{Abdi2010a} {Abdi, H., and Williams, L.J. (2010) {\it
Principal component analysis}. Wiley, New York.}

\bibitem{Abdi2010b} {Abdi, H. (2010) Partial least square
regression, projection on latent structure regression,
PLS-Regression. {\it Computational Statistics} {bf 2}: 97-106.}

\bibitem{Ahlfors2010} {Ahlfors SP, Han J, Lin FH, Witzel T, 
Belliveau JW, H\"am\"al\"ainen MS and Halgren E. (2010) Cancellation of EEG 
and MEG signals generated by extended and distributed sources. {\it 
Hum. Brain Mapping} {\bf 31}: 140-149.}

\bibitem{Bed2004} B\'edard, C., Kr\"oger, H. and Destexhe, A. (2004)
Modeling extracellular field potentials and the frequency-filtering
properties of extracellular space. {\it Biophys. J.} {\bf 86}:
1829-1842.

\bibitem{Bed2006a} B\'edard, C., Kr\"oger, H. and Destexhe, A. (2006a) Does
  the $1/f$ frequency-scaling of brain signals reflect self-organized critical
  states~? {\it Physical Review Letters} {\bf 97}: 118102.

\bibitem{Bed2006b} {B\'edard, C., Kr\"oger, H., Destexhe, A. (2006b)
    Model of low-pass filtering of local field potentials in brain tissue.
    {\it Phys. Rev. E}~{\bf 73}: 051911.}

\bibitem{Bed2009} B\'edard, C. and Destexhe, A. (2009) Macroscopic
models of local field potentials and the apparent $1/f$ noise in
brain activity. {\it Biophys. J.} {\bf 96}: 2589-2603.

\bibitem{TransferFct2009} B\'edard, C., Rodrigues, S., Roy, N.,
Contreras, D. and Destexhe, A. (2010) Evidence for
frequency-dependent extracellular impedance from the transfer
function between extracellular and intracellular potentials.  {\it
J. Computational Neurosci.}, in press.

\bibitem{Bell-Sejnowski} Bell A.J. and Sejnowski T.J. (1995) An
information maximisation approach to blind separation and blind
deconvolution. {\it Neural Computation} {\bf 7}: 1129-1159.

\bibitem{Berouti79} {Berouti M, Schwartz R and Makhoul J.
(1979) Enhancement of speech corrupted by acoustic noise.  {\it
Proc. ICASSP 1979}, 208-211.}

\bibitem{Boll79} {Boll S.F. (1979) Suppression of acoustic noise in 
speech using spectral subtraction.  {\it IEEE Trans. Acoustic, Speech and 
Signal Processing} {\bf 27}: 113-120.}

\bibitem{Boubakir2007} {Boubakir C., Berkani D. and Grenez
F. (2007) A frequency-dependent speech enhancement method. {\it J.
Mobile Communication} {\bf 1}: 97-100.}

\bibitem{Buiatti2007} Buiatti M, Papo D, Baudonni\`ere PM
and van Vreeswijk C. (2007) Feedback modulates the temporal
scale-free dynamics of brain electrical activity in a hypothesis
testing task.  {\it Neuroscience} {\bf 146}: 1400-1412. 

\bibitem{Buzsaki2004} {Buzs\'aki G and Draguhn A. (2004) Neuronal
oscillations in cortical networks. {\it Science} {\bf 304}: 1926-1929.}

\bibitem{Cuffin79} {Cuffin BN and Cohen D. (1979) Comparison of the
    magnetoencephalogram and electroencephalogram. {\it EEG
      Clin. Neurophysiol.} {bf 47}: 132-146.}

\bibitem{deBoor2001} de Boor, C. (2001) {\it A Practical Guide to
Splines}, Springer-Verlag, New York, revised edition, 2001.

\bibitem{Delorme2004} Delorme, A. and Makeig, S. (2004) EEGLAB: an
open source toolbox for analysis of single-trial EEG dynamics
including independent component analysis. {\it J. Neurosci.
Methods} {\bf 134}: 9-21.

\bibitem{Des2003} {Destexhe, A., Rudolph, M. and Par\'e, D. (2003) The
    high-conductance state of neocortical neurons {\it in vivo}.  {\it Nature
      Reviews Neurosci.} {\bf 4}: 739-751.}

\bibitem{Diard1999} {Diard, J-P., Le Gorrec, B. and Montella, C. (1999)
    Linear diffusion impedance. General expression and applications.  {\it
      J. Electroanalytical Chem.  }. {\bf 471}: 126-131.}

\bibitem{Eilers1996} Eilers, P.H.C. and Marx, B.D. (1996) Flexible
smoothing with B-splines and penalties. {\it Statistical Science}
{\bf 11}: 89–121.

\bibitem{Ferree2001} {Ferree TC, Luu P, Russell GS and Tucker 
DM. (2001) Scalp electrode impedance, infection risk, and EEG data 
quality. {\it Clin. Neurophysiol.} \bf{112}: 536-544.}

\bibitem{Foster1989} Foster, KR. and Schwan, HP. (1989)
Dielectric properties of tissues and biological materials: a
critical review.  {\it Crit. Reviews Biomed. Engineering} {\bf 17}:
25-104.

\bibitem{Freeman2000} Freeman WJ, Rogers LJ, Holmes MD and Silbergeld
DL. (2000) Spatial spectral analysis of human electrocorticograms
including the alpha and gamma bands. {\it J. Neurosci. Methods} {\bf
95}: 111-121.

\bibitem{Gabriel1996a} Gabriel, S., Lau, R.W. and Gabriel, C.
(1996a) The dielectric properties of biological tissues : I.
Literature survey. {\it Phys. Med. Biol.}. {\bf 41 }: 2231-2249.

\bibitem{Gabriel1996b} Gabriel, S., Lau, R.W. and Gabriel, C.
(1996b) The dielectric properties of biological tissues : II.
Measurements in the frequency range 10 Hz to 20 GHz. {\it Phys. Med.
Biol.}. {\bf 41}: 2251-2269.

\bibitem{Gabriel1996c} Gabriel, S., Lau, R.W. and Gabriel, C.
(1996c) The dielectric properties of biological tissues : III.
Parametric models for the dielectric spectrum tissues. {\it Phys.
Med. Biol.}.  {\bf 41 }: 2271-2293.

\bibitem{Garthwaite94} {Garthwaite P. (1994) An interpretation of
partial least squares. {\it J. Am. Stat. Assoc.} {\bf 89}: 122-127.}

\bibitem{Gulrajani98} Gulrajani, R.M. (1998) {\it
Bioelectricity and Biomagnetism}, Wiley, New York.

\bibitem{Hamalainen93} H\"am\"al\"ainen, M., Hari, R., Ilmoniemi,
R.J., Knuutila, J. and Lounasmaa, O.V. (1993) Magnetoencephalography
-- theory, instrumentation, and applications to noninvasive studies
of the working human brain. {\it Reviews Modern Physics} {\bf 65}:
413-497.

\bibitem{Hwa2002a} {Hwa RC and Ferree TC (2002a). Scaling 
properties of fluctuations in the human electroencephalogram. {\it 
Phys Rev E} {\bf 66}: 021901.}

\bibitem{Hwa2002b} {Hwa RC and Ferree TC (2002b). Fluctuation 
analysis of human electroencephalogram.  {\it Nonlinear Phenomena
in Complex Systems} {\bf 5}: 302-307.}

\bibitem{Kamath2002} {Kamath, S. and Loizou, P. (2002). A
multi-band spectral subtraction method for enhancing speech
corrupted by colored noise.  {\it Proceedings of ICASSP 2002},
4160-4164.}

\bibitem{Kantelhardt2001} {Kantelhardt J, Koscielny-Bunde E, 
Rego H, Havlin S and Bunde A. (2001) Detecting long-range correlations 
with detrended fluctuation analysis. {\it Phys A} {\bf 295}: 441-454. }

\bibitem{Katkovnik2006} Katkovnik,V., Egiazarian, K. and
Astola, J. (2006) Local approximation in signal and image
processing, SPIE Press. 

\bibitem{Landau84} Landau L and Lifchitz E. (1984) {\it
Electrodynamics of Continuous Media}, MIR Editions, Moskow.

\bibitem{Lim79} {Lim, J.S. and Oppenheim, A.V. (1979) Enhancement
and band width compression of noisy speech.  {\it Proc. of the IEEE}
{\bf 67}: 1586-1604.}

\bibitem{Linkenkaer-Hansen2001} {Linkenkaer-Hansen K, Nikouline VV,
Palva JM, and Ilmoniemi RJ. (2001) Long-range temporal correlations
and scaling behavior in human brain oscillations. {\it J.
Neurosci.} {\bf 21}: 1370-1377.}

\bibitem{Logo2007} Logothetis, N.K., Kayser, C. and Oeltermann, A.
(2007) In vivo measurement of cortical impedance spectrum in monkeys:
Implications for signal propagation.  {\it Neuron} {\bf 55}: 809-823.

\bibitem{Loizou2007} Loizou, PC. (2007) {\it Speech
Enhancement: Theory and Practice}, CRC Press, Boca Raton: FL.

\bibitem{Magee98} Magee L. (1998) Nonlocal behavior
in polynomical regression. {\it The American
Statistican} {\bf 52}: 20-22.

\bibitem{Miller2009} {Miller KJ, Sorensen LB, Ojemann JG, and den Nijs M. 
(2009) Power-law scaling in the brain surface electric potential. {\it PLoS 
Comput Biol.} {\bf 5}: e1000609.}

\bibitem{Nenonen} Nenonen, J., Kajola, M., Simola, J. and Ahonen, A.
(2004) Total information of multichannel MEG sensor arrays. In: {\it
Proceedings of the 14th International Conference on Biomagnetism
(Biomag2004)}, edited by Halgren, E., Ahlfors, A., Hamalainen, M. and
Cohen, D., Boston, MA, pp.~630–631.

\bibitem{Novikov97} Novikov, E., Novikov, A., Shannahoff-Khalsa, D.,
Schwartz, B. and Wright, J. (1997) Scale-similar activity in the
brain. {\it Phys. Rev. E} {\bf 56}: R2387-R2389.

\bibitem{Nunez2005} Nunez, P.L. and Srinivasan, R. (2005) {\it
Electric Fields of the Brain. The Neurophysics of EEG} (2nd edition).
Oxford university press, Oxford, UK.

\bibitem{Plonsey69} Plonsey, R. (1969) {\it Bioelectric
Phenomena}, McGraw Hill, New York.

\bibitem{Pritchard92} Pritchard, W.S. (1992) The brain in fractal
time: $1/f$-like power spectrum scaling of the human
electroencephalogram.  {\it Int. J. Neurosci.} {\bf 66}: 119-129.

\bibitem{Rall68} Rall, W. and Shepherd, G.M. (1968)  Theoretical
reconstruction of field potentials and dendrodendritic synaptic
interactions in olfactory bulb. {\it J. Neurophysiol.} {\bf 31}:
884-915.

\bibitem{Ramirez2008} {Ramirez RR. (2008) Source localization.
{\it Scholarpedia} {\bf 3} 1733.}

\bibitem{Ranck63} Ranck, J.B., Jr. (1963) Specific impedance of
rabbit cerebral cortex. {\it Exp. Neurol.} {\bf 7}: 144-152.

\bibitem{Royston94} Royston P and Altman D. (1994)
Regression using fractional polynomials of continuous covariates:
parsimonious parametric modelling. {\it Applied Statistician} {\bf
43}: 429-467.

\bibitem{Sarvas87} {Sarvas, L. (1987) Basic mathematical and 
electromagnetic concepts of the biomagnetic inverse problem.
{\it Phys. Med. Biol.} {\bf 32}: 11-22.}

\bibitem{Sim98} {Sim BL, Tong YC, Chang JC and Tan CT. (1998).  A
Parametric formulation of the generalized spectral subtraction
method. {\it IEEE Trans. Speech and Audio Processing} {\bf 6}:
328-337.}

\bibitem{Tay1995} {Taylor, S.R. and Gileadi, E. (1995) The physical
interpretation of the Warburg impedance. {\it Corrosion} {\bf 51}:
664-671.}

\bibitem{Valencia2008} {Valencia M, Artieda J, Alegre M, and 
Maza D. (2008) Influence of filters in the detrended fluctuation 
analysis of digital electroencephalographic data. {\it J. Neurosci. 
Methods} {\bf 170}: 310-316.}

\bibitem{Voipio2003} {Voipio J, Tallgren P, Heinonen E, Vanhatalo
S, and Kaila K. (1989) Millivolt-scale DC shifts in the human scalp
EEG: evidence for a nonneuronal generator. {\it J. Neurophysiol.}
{\bf 89}: 2208-2214.}

\bibitem{Wolters2007} {Wolters, C and de Munck JC.  (2007) Volume
conduction. {\it Scholarpedia} {\bf 2}: 1738.}

\end{description}

%\clearpage

\section*{{Supplementary material}}

\subsection*{{Supplementary methods}}

{We give details below to some of the methods and quantities
used in the Results.}  

{\subsubsection*{SNR}}

{Two of the used methods for noise-correction are based on
band-specific signal-to-noise ratio (SNR) in order to cancel the
effects of background colored-noise in the spectra of interest. In
each subject, average PSD was used to calculate signal-to-noise
ratio (SNR). For SNR calculation, few frequency bands were defined
based on the categorization in Buzsaki \& Draguhn (2004): 0-10 Hz
(Slow, Delta and Theta), 11-30 Hz (Beta), 30-80 Hz (Gamma), 80-200
Hz (Fast oscillation), 200-500 Hz (Ultra-fast oscillation). SNR was
calculated as:
\begin{equation}
	SNR_{bi} =
\frac{\sum{10*log10(\frac{PSDsignal_{bi}}{PSDnoise_{bi}})}}{n} 
\label{SNR}
\end{equation}
for a given band "b" and sensor "i", "n" is the frequency
resolution of that band. This method was applied on individual
average PSD as well as shape preserving spline of each average PSD
where each PSD was fist smoothed in log10 scale using a shape
preserving spline, i.e, Piecewise Cubic Hermite Interpolating
Polynomial (PCHIP).}

\subsubsection*{{Multiband spectral subtraction}}

{Assuming the additive noise to be stationary and uncorrelated with
the clean signal, nearly most spectral subtraction methods can be
formulated using a parametric equation: 
 \begin{equation}
\lvert\widehat{S(k)}\rvert^{\alpha} = a_{k}\lvert Y(k)
\rvert^{\alpha} - b_{k}\lvert\widehat{D(k)}\rvert^{\alpha}
\label{LMSS}
\end{equation}
where $\lvert\widehat{{S}}_k\rvert$, $\lvert{Y}_k\rvert$ and
$\lvert\widehat{{D}}_k\rvert$ refer to enhanced magnitude spectrum
estimate (corrected signal), the noisy magnitude spectrum (original
signal) and noise magnitude spectrum estimate (``noise''),
respectively.  $k$ is the frequency index, while $a_{k}$ and
$b_{k}$ are linear coefficient parameters of the summation. 
Spectral subtraction methods fall into three main categories (Sim
et al., 1998). The simplest of all, a linear method where $a_{k}$ =
$b_{k}$ = 1, $\alpha$=2, following Boll et al.\ (1979) was used
here.  This linear multiband spectral subtraction (LMSS) method is
well-established for noise subtraction (see Loizou, 2007 for a
comparative study of noise subtraction methods).}

{An improved method, with $a_{k}$ = 1 and $b_{k}$ = v, where
"v" is the oversubtraction factor. This method uses oversubtraction
and introduces a spectral flooring to minimize residual noise and
musical noise (Berouti et al., 1979).  A second category of
spectral subtraction is based on $a_{k}$ = $b_{k}$ = \textit{f(k)}.
Third and the most robust methods are based on a non-linear
multiband subtraction (NMSS) where $a_{k}$ = 1 and $b_{k}$ =
\textit{v(k)}; i.e., the oversubtraction factor is adjusted based
on a specfic band's SNR. These methods proposed by (Kamath and
Loizou, 2002; Loizou, 2007) are suitable for dealing with colored noise
(Boubakir et al., 2007; Sim et al., 1998), a case similar to MEG
recordings. The spectrum is divided into N non-overlapping bands,
and spectral subtraction is performed independently in each band.
The Eqs.~\ref{LMSS} is simply reduced to:
\begin{equation}
 	\lvert\widehat{{S}_i(k)}\rvert^{2} = \lvert{Y}_i(k)\rvert^{2} -
{\alpha}_i {\delta}_i \lvert\widehat{{D}_i(k)}\rvert^{2}, b_{i}\leq k \leq e_{i}
\label{NMSS}
\end{equation}
where $b_{i}$ and $e_{i}$ are the beginning and ending frequency
bins of the ith frequency band, $\alpha_{i}$ is the overall
oversubtraction factor of the ith band and $\delta_{i}$ is a
tweaking factor. The band specific oversubtraction factor
$\alpha_{i}$ is a function of the segmental $SNR_{i}$ of the ith
frequency band. After calculating bandspecific SNR
(Eqs.~\ref{SNR}), we used the product of lower 10 percent of
crosssubject average SNR and standard deviation of $SNR_{i}$ to
estimate the $\alpha_{i}$ $\delta_{i}$ subtraction coefficient. 
Next, simply by multiplying the noise PSD by this coefficient and
subtracting it from the measured PSD, the enhanced PSD was
achieved.}

{\subsubsection*{Wiener filter (WF) spectral enhancement}}

{The principle of the Wiener filter is to obtain an estimate
of the clean signal from that of the noisy measurement through
minimizing the Mean Square Error (MSE) between the desired and the
measured signal (Lim et al., 1979; Abd El-Fattah et al., 2008). In
the frequency domain, this relation is formulated as filtering
transfer function:
\begin{equation}
 	WF(k) = \frac{P_{s}(k)}{P_{s}(k) + P_{n}(k)}
\label{WFfrq}
\end{equation}
where, as before, $P_{s}(k)$ and $P_{n}(k)$ refer to enhanced power
spectrum estimate and noise power spectrum estimate respectively
for a signal frame and $k$ is the frequency index. Based on the
definition of SNR as, the ratio of these two elements, one can
formulate the WF as:
\begin{equation}
	WF_{K} = [1+\frac{1}{SNR_{k}}]^{-1}
\label{WFsnr}
\end{equation}
After calculation of bandspecific WF, the noisy signal is simply
muliplied by the WF to obtain the enhanced signal.}

{\subsubsection*{Partial least square (PLS) approximation of
non-noisy spectrum}}

{Partial least squares (PLS) regression, combines ``Principal
component analysis" (PCA) and ``Multiple linear regression" (Abdi,
2010; Abdi and Williams, 2010). While PCA finds hyperplanes of
maximum variance between the response and independent variables,
PLS projects the predicted variables and the observable variables
to a new space. Then from this new space, it finds a linear
regression model for the projected data. Next, using this model,
PLS finds the multidimensional direction in the X space that
explains the maximum multidimensional variance direction in the Y
space (Abdi, 2010; Garthwaite, 1994). If X is the PSD of noise
measurement and Y is the PSD of the measured signal contaminated
with background noise, one can use PLS to "clean" one matrix (Y) by
predicting Y from X and then using the residual of the prediction
of Y by X as the estimate of pure PSD. The patterns of the awake
spectrum that statistically resembles the patterns of emptyroom
spectral noise are those that should be removed. As during PLS
algorithm, the data is mean subtracted and z-normalized, the
predection of Y from X is an approximate of the zscored PSD. 
Therefore, the reseidual Y, which is taken as the spectral features
that can not be predicted by noise, also has zscored values. It has
too be emphasized that this approach of denoising only works in the
spectral but not the time domain.}

%\clearpage
\subsection*{{Supplementary table}}

%-----------------------------------------------
%    Table
%-----------------------------------------------

\begin{table}[h!]

\ \\ {\bf A. Mean and standard deviation}

%{\small
%\begin{tabular}{c|llllllll}
%       & EEG              & MEG (awake)      & MEG(empty)       & LMSS             & NMSS             & WF               & PLS              & ES  \\
%\hline
%All    & -1.33 $\pm$ 0.19 & -1.24 $\pm$ 0.26 & -1.04 $\pm$ 0.13 & -1.24 $\pm$ 0.28 & -1.06 $\pm$ 0.29 & -1.05 $\pm$ 0.27 & -0.50 $\pm$ 0.11 & -0.20 $\pm$ 0.23 \\
%FR ROI & -1.36 $\pm$ 0.25 & -0.97 $\pm$ 0.10 & -0.97 $\pm$ 0.06 & -0.96 $\pm$ 0.11 & -0.76 $\pm$ 0.09 & -0.76 $\pm$ 0.08 & -0.40 $\pm$ 0.05 & -0.00 $\pm$ 0.09 \\
%VX ROI & -1.21 $\pm$ 0.13 & -1.36 $\pm$ 0.10 & -1.10 $\pm$ 0.09 & -1.36 $\pm$ 0.10 & -1.14 $\pm$ 0.11 & -1.12 $\pm$ 0.11 & -0.50 $\pm$ 0.04 & -0.26 $\pm$ 0.08 \\
%PT ROI & -1.36 $\pm$ 0.12 & -1.30 $\pm$ 0.29 & -1.08 $\pm$ 0.15 & -1.31 $\pm$ 0.32 & -1.16 $\pm$ 0.32 & -1.14 $\pm$ 0.30 & -0.54 $\pm$ 0.11 & -0.22 $\pm$ 0.26 \\
%\hline
%\end{tabular}
%}

{\small
\begin{tabular}{c|llll}
       & EEG              & MEG (awake)      & MEG(empty)       & LMSS \\
\hline
All    & -1.33 $\pm$ 0.19 & -1.24 $\pm$ 0.26 & -1.04 $\pm$ 0.13 & -1.24 $\pm$ 0.28 \\
FR ROI & -1.36 $\pm$ 0.25 & -0.97 $\pm$ 0.10 & -0.97 $\pm$ 0.06 & -0.96 $\pm$ 0.11 \\
VX ROI & -1.21 $\pm$ 0.13 & -1.36 $\pm$ 0.10 & -1.10 $\pm$ 0.09 & -1.36 $\pm$ 0.10 \\
PT ROI & -1.36 $\pm$ 0.12 & -1.30 $\pm$ 0.29 & -1.08 $\pm$ 0.15 & -1.31 $\pm$ 0.32 \\
\hline
\end{tabular}
}

\

{\small
\begin{tabular}{c|llll}
       & NMSS             & WF               & PLS              & ES  \\
\hline
All    & -1.06 $\pm$ 0.29 & -1.05 $\pm$ 0.27 & -0.50 $\pm$ 0.11 & -0.20 $\pm$ 0.23 \\
FR ROI & -0.76 $\pm$ 0.09 & -0.76 $\pm$ 0.08 & -0.40 $\pm$ 0.05 & -0.00 $\pm$ 0.09 \\
VX ROI & -1.14 $\pm$ 0.11 & -1.12 $\pm$ 0.11 & -0.50 $\pm$ 0.04 & -0.26 $\pm$ 0.08 \\
PT ROI & -1.16 $\pm$ 0.32 & -1.14 $\pm$ 0.30 & -0.54 $\pm$ 0.11 & -0.22 $\pm$ 0.26 \\
\hline
\end{tabular}
}

\ \\ {\bf B. Pearson correlation of EEG vs.}

{
\begin{tabular}{c|llllll}
       & MEG & LMSS & NMSS & WF & PLS & ES \\
All    & 0.29  &   0.29  &   0.32   &  0.33  &   0.37  &  0.35  \\
FR ROI & 0.41  &   0.39  &   0.32   &  0.37  &   0.01  &  0.17  \\
VX ROI & -0.17 &   -0.10 &  -0.15   &  -0.13 &   0.01  &  -0.28 \\
PT ROI & 0.35  &   0.34  &   0.38   &  0.39  &  0.46  &  0.41 \\
\hline
\end{tabular}
}

\ \\ {\bf C. Kendall Rank Corr of EEG vs.}

{
\begin{tabular}{c|llllll}
       & MEG & LMSS & NMSS & WF & PLS & ES \\
\hline
All    &  0.21  &  0.21  &  0.24   &  0.25  &   0.29  &   0.23\\
FR ROI &  0.29  &  0.23  &  0.21   &  0.27  &   -0.06 &   0.12\\
VX ROI &  -0.03 &  0.04  &  -0.04  &  -0.03 &   0.07  &   -0.09\\
PT ROI &  0.23  &  0.23  &  0.26   &  0.26  &   0.30  &  0.27\\
\hline
\end{tabular}
}

\caption{{ROI statistical comparison for different noise
correction methods. A. mean and std of frequency scale exponent for
all regions and individual ROI. B.  numerical values of linear
Pearson correlation. C. rank-based Kendall correlation.}}

\label{supplTable}
\end{table}
%-----------------------------------------------

\subsection*{{Supplementary figures}}

%-----------------------------------------------
%    Figure
%-----------------------------------------------
\begin{figure}[h!]
\includegraphics[width=\columnwidth]{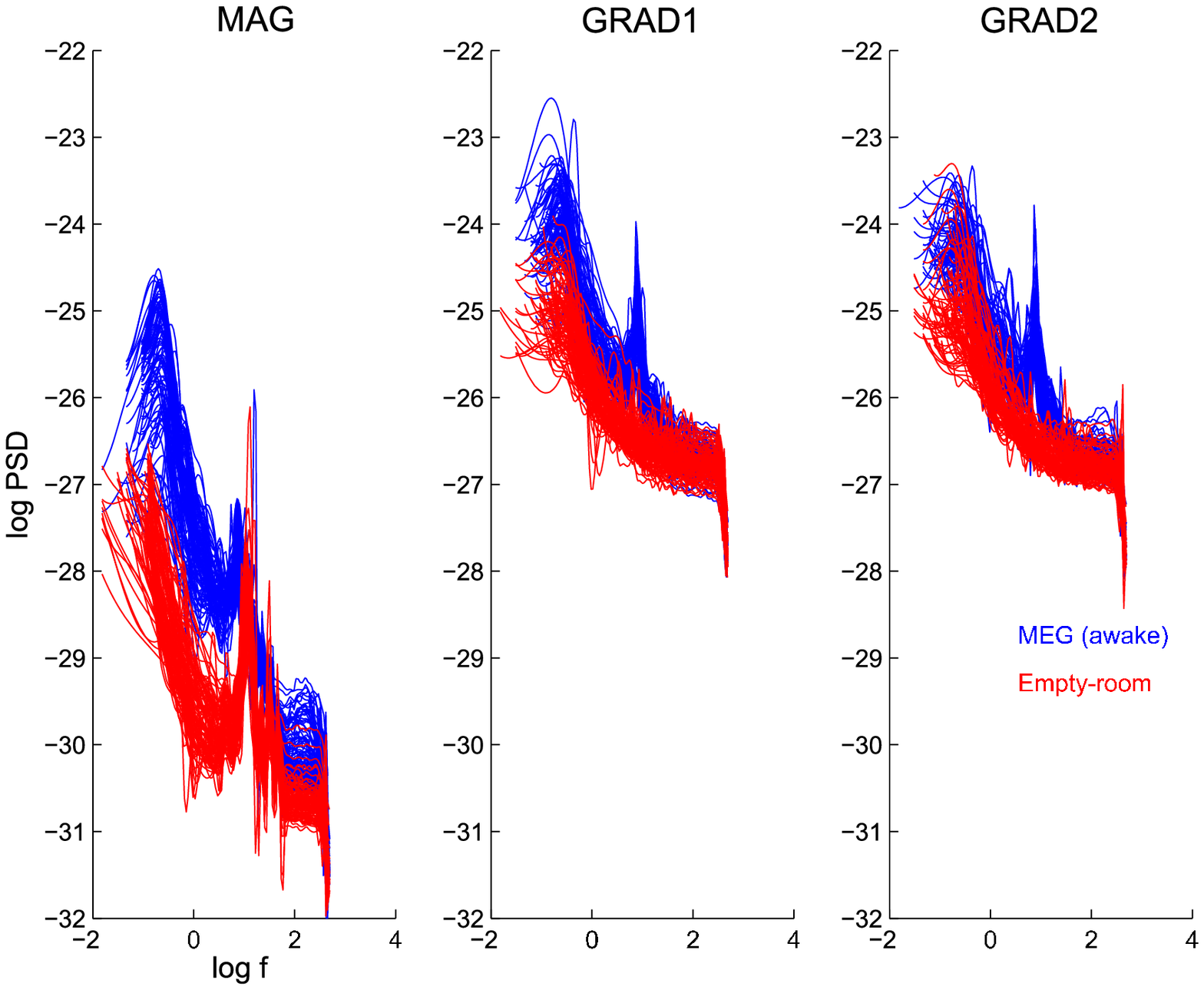}

{Figure S1:  Frequency spectra of magnetometers and
gradiometers.  Comparison of awake (blue) vs empty-room (red)
recordings between Magnetometers (MAG) and Gradiometers (GRAD1,
GRAD2) in a sample subject.  As for the EEG, the MEG signal is
characterized by a peak at around 10~Hz, which is presumably due to
residual alpha rhythm (although the subject had eyes open).  This
is also visible from the MEG signals (Fig.~\ref{EEGMEG}) as well as
from their PSD (Fig.~\ref{PSD} and MAG panel here).  The power
spectrum from the empty-room signals also show a peak at around
10~Hz, but this peak disappears from the gradiometer empty-room
signals, while the 10~Hz peak of MEG still persists for
gradiometers awake recordings.  This suggests that these two 10~Hz peaks are
different oscillation phenomena.  All other subjects showed a
similar pattern.}

\label{Noise}
\end{figure}
%-----------------------------------------------

%-----------------------------------------------
%    Figure
%----------------------------------------------- 
\begin{figure}[h!]
%\centering 
\includegraphics[width=\columnwidth]{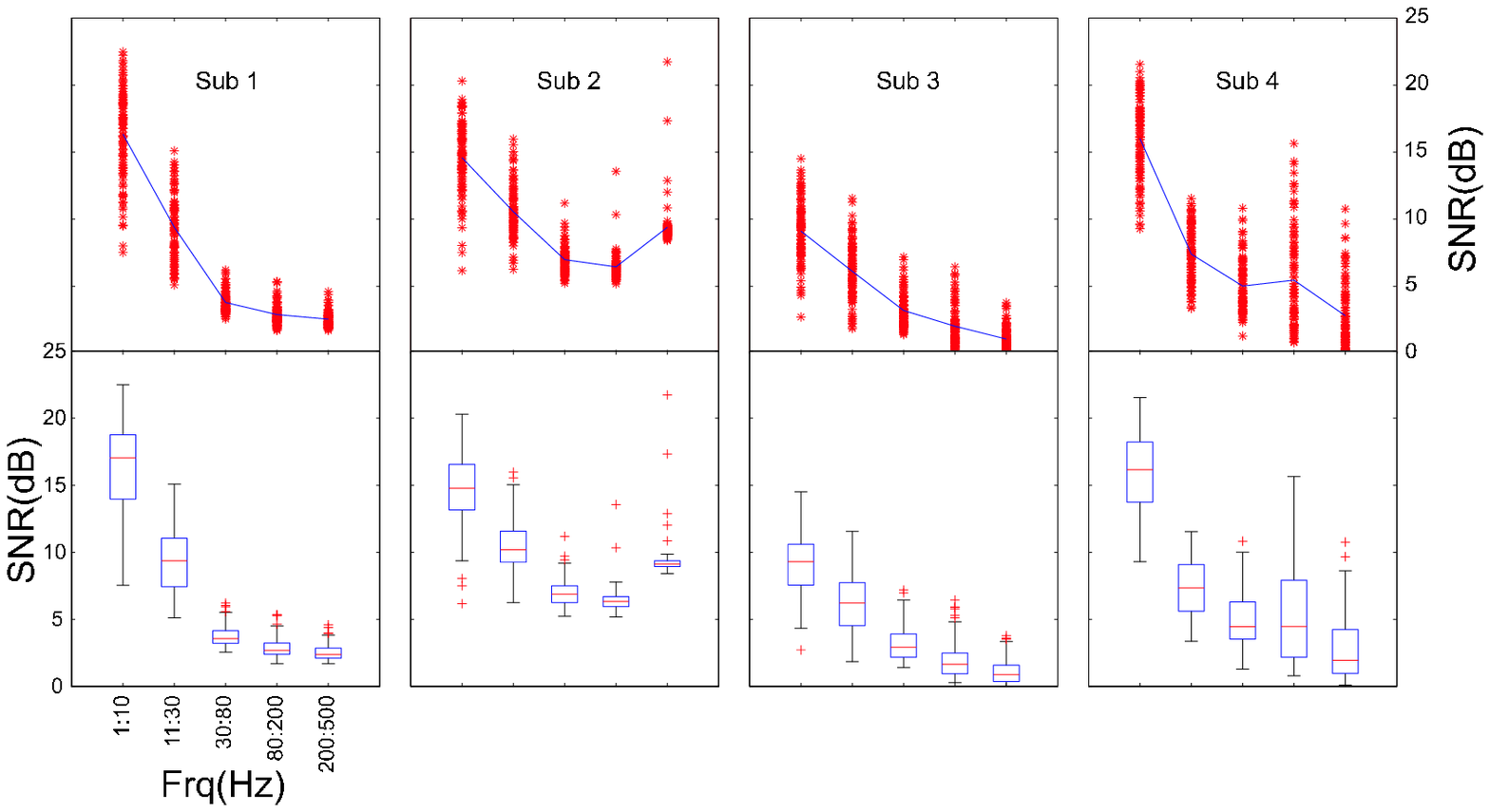}

{Figure S2: Signal-to-noise ratio (SNR) of Magnetometers (MAG)
for multiple frequency bands: 0-10 Hz (Slow, Delta and Theta),
11-30 Hz (Beta), 30-80 Hz (Gamma), 80-200 Hz (Fast oscillation),
200-500 Hz (Ultra-fast oscillation). In the scatterplots, red
astrisks relate to individual sensors and the blue line is the
band-specific mean across the sensors. In boxplots, the box has
lines at the lower quartile, median (red), and upper quartile
values. Smallest and biggest non-outlier observations (1.5 times
the interquartile range IRQ) are shown as whiskers.  Outliers are
data with values beyond the ends of the whiskers and are displayed
with a red + sign. In all subjects, the SNR shows a band-specific
trend and has the highest value for lower frequencies and gradually
drops down as band frequency goes up. As the frequency drops, the
variability of SNR (among sensors) rises; therefore, the SNR of the
lowest band (1-10~Hz) shows the highest sensors-to-sensor
variability and the highest SNR in comparison to other frequency
bands.}

\label{SNRchart}
\end{figure}
%-----------------------------------------------

%-----------------------------------------------
%    Figure
%----------------------------------------------- 
\begin{figure}[h!]
%\centering 
\includegraphics[width=\columnwidth]{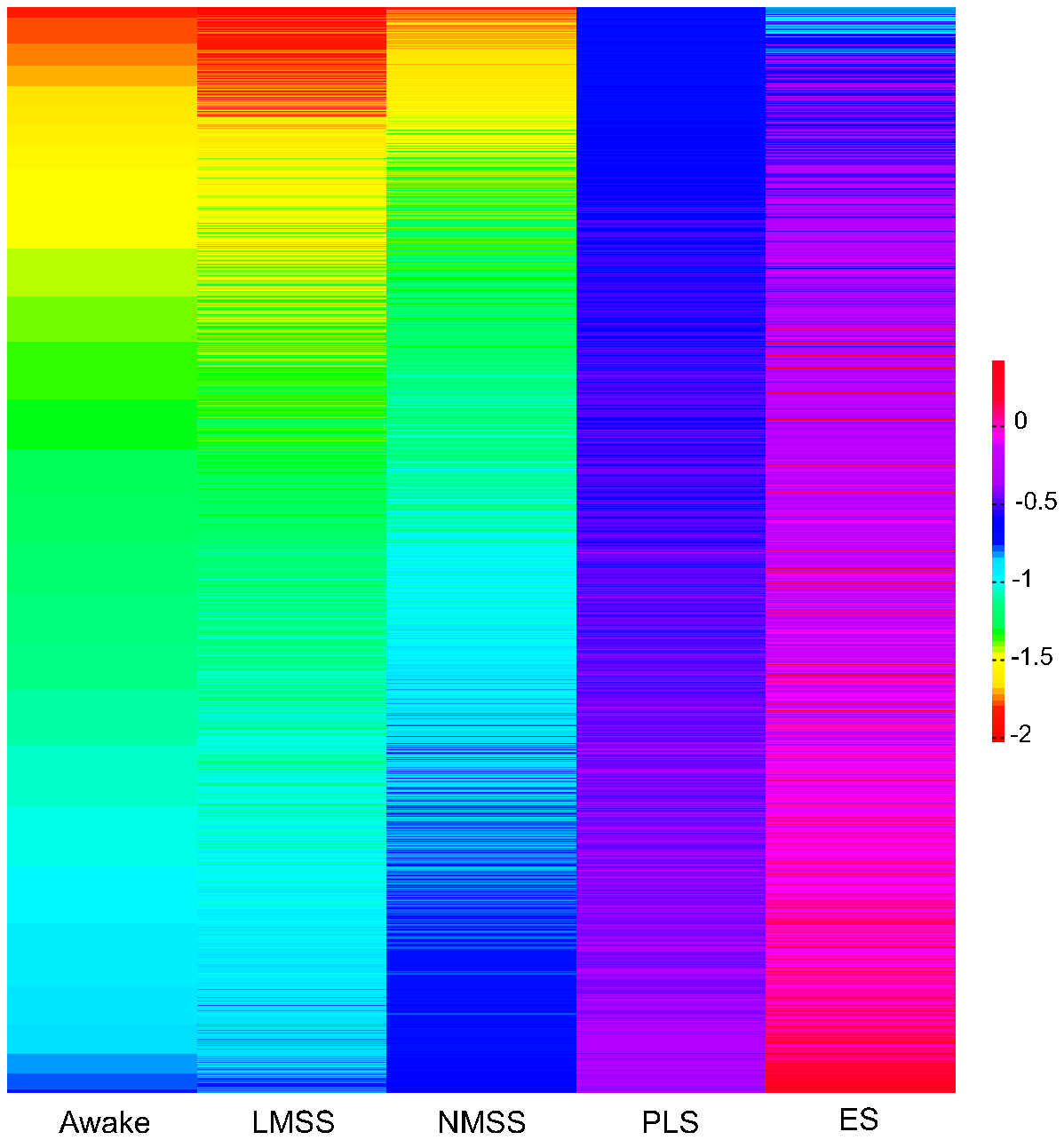}

{Figure S3: Noise correction comparison. Every horizontal line
showes a voxel of the topographical maps shown in Fig. \ref{topo}
sorted based on the scaling exponent values of awake MEG (left
stripe). Using a continuous color spectrum, these stripes show that
minimal correction is achived by LMSS. As indicated in the text,
the performance of this method is not reliable due to the nonlinear
nature of SNR (see Suppl.\ Fig.~S2). NMSS yields higher degree of
correction. WF performs almost identical to NMSS (not shown here). 
Exponent subtraction almost abolishes the sacling all together (far
right stripe). PLS results in values between NMSS and "Exponent
subtraction". For details of each of these correction procedures,
see Methods. LMSS, NMSS and WF rely on additive uncorrelated nature
of noise.  ``Exponent subtraction" assumes that the noise is
intrinsic to SQUID. PLS ascertains the characteristics of noise to
the collective obeserved pattern of spectral domain across all
frequencies. See text for more details.}

\label{Correction}
\end{figure}
%-----------------------------------------------

\end{document}